%% file: paper.tex
\documentclass[acmtog,timestamp]{acmart}
\usepackage{booktabs}           % For formal tables
\usepackage{subcaption}         % for subcaption
\usepackage{overpic}

\usepackage[normalem]{ulem}     % for marking strike through
\usepackage{color}
\definecolor{todoCol}{rgb}{0.9, 0, 0}

\definecolor{meCol}{rgb}{0, 0, 0}
\definecolor{kuCol}{rgb}{0, 0, 0}
\definecolor{ntCol}{rgb}{0, 0, 0.}

\newcommand{\ku}[1]{{\color{kuCol}#1}}
\newcommand{\nt}[1]{{\color{ntCol}#1}}
\newcommand{\nils}[1]{{\color{ntCol}#1}} % helper
\newcommand{\me}[1]{{\color{meCol}#1}}

\usepackage[utf8]{inputenc}
\usepackage{mathtools}
\usepackage{algorithm}
\usepackage[noend]{algpseudocode}
\MakeRobust{\Call}

 \setcopyright{acmlicensed}
\usepackage{wrapfig}

\newcommand{\x}{\mathbf{x}}
\newcommand{\y}{\mathbf{y}}
\newcommand{\z}{\mathbf{z}}
\newcommand{\xV}{\mathbf{x}_\mathbf{u}}
\newcommand{\yV}{\mathbf{y}_\mathbf{u}}
\newcommand{\zV}{\mathbf{z}_\mathbf{u}}
\newcommand{\xD}{\mathbf{x}_{\Phi}}
\newcommand{\yD}{\mathbf{y}_{\Phi}}
\newcommand{\zD}{\mathbf{z}_{\Phi}}
\newcommand{\PP}{\mathbf{P}}
\newcommand{\R}{\mathbf{R}}
\newcommand{\PPP}{\mathbf{P}^T\mathbf{P}}

\newcommand{\norm}[1]{\left\lVert#1\right\rVert}

\newcommand{\half}{\tfrac{1}{2}}
\newcommand{\Phy}[1]{\Phi^{#1}}
\newcommand{\PhiI}[1]{\Phi^{#1}_{\text{I}}}
\newcommand{\PhiIp}[1]{\tilde\Phi^{#1}_{\text{I}}}
\newcommand{\PhiIpn}[1]{\tilde\Phi^{#1}_{\text{I,new}}}

\newcommand{\PhyNoT}{\Phi}
\newcommand{\Phip}[1]{\tilde\Phi^{#1}}
\newcommand{\Phiu}[1]{\Delta\Phi^{#1}}
\newcommand{\Phic}[1]{\Delta\Phi^{#1}_c}
\renewcommand{\u}[1]{\mathbf{u}^{#1}}
\newcommand{\uNoT}{\mathbf{u}}
\newcommand{\up}[1]{\mathbf{\tilde u}^{#1}}
\newcommand{\uu}[1]{\Delta\u{#1}}
\newcommand{\uuI}[1]{\Delta\u{#1}_{\text{tmp}}}
\newcommand{\uuC}[1]{\Delta\u{#1}_{\text{C}}}

\newcommand{\fNoT}{\mathbf{f}_{\text{ext}}}
\renewcommand{\i}[1]{i^{#1}}
\newcommand{\iNoT}{i}
\newcommand{\ip}[1]{\tilde i^{#1}}
\newcommand{\iu}[1]{\Delta i^{#1}}

\newcommand{\myrefeq}[1]{Eq.~\eqref{#1}}
\newcommand{\myreffig}[1]{Fig.~\ref{#1}}

\newcommand{\myrefsec}[1]{Sec.~\ref{#1}}
\newcommand{\myrefapp}[1]{Appendix~\ref{#1}}
\newcommand{\myrefalg}[1]{Alg.~\ref{#1}}

\newcommand{\acro}{ScalarFlow}

\begin{document}
% Title portion
\title{\acro{}: A Large-Scale Volumetric Data Set of Real-world Scalar Transport Flows for Computer Animation and Machine Learning}

 \author{Marie-Lena Eckert}
 \affiliation{%
 \institution{Technical University of Munich}
 \streetaddress{Boltzmannstr. 3}
 \city{Garching b. Munich}
 \postcode{85748}
 \country{Germany}}
 \email{marie-lena.eckert@tum.de}

\author{Kiwon Um}
\affiliation{%
	\institution{Technical University of Munich}
	\streetaddress{Boltzmannstr. 3}
	\city{Garching b. Munich}
	\postcode{85748}
	\country{Germany}}
\email{kiwon.um@tum.de}

\author{Nils Thuerey}
\affiliation{%
	\institution{Technical University of Munich}
	\streetaddress{Boltzmannstr. 3}
	\city{Garching b. Munich}
	\postcode{85748}
	\country{Germany}}
\email{nils.thuerey@tum.de}

\begin{abstract}
  \input{0_abstract}
\end{abstract}

%
% The code below should be generated by the tool at
% http://dl.acm.org/ccs.cfm
% Please copy and paste the code instead of the example below.
%
% TODO adapt


%
% End generated code
%

\acmJournal{TOG}
\acmYear{2019}\acmVolume{38}\acmNumber{6}\acmArticle{239}\acmMonth{11} \acmDOI{10.1145/3355089.3356545}


\begin{teaserfigure}
  \centering
  \includegraphics[width=\linewidth]{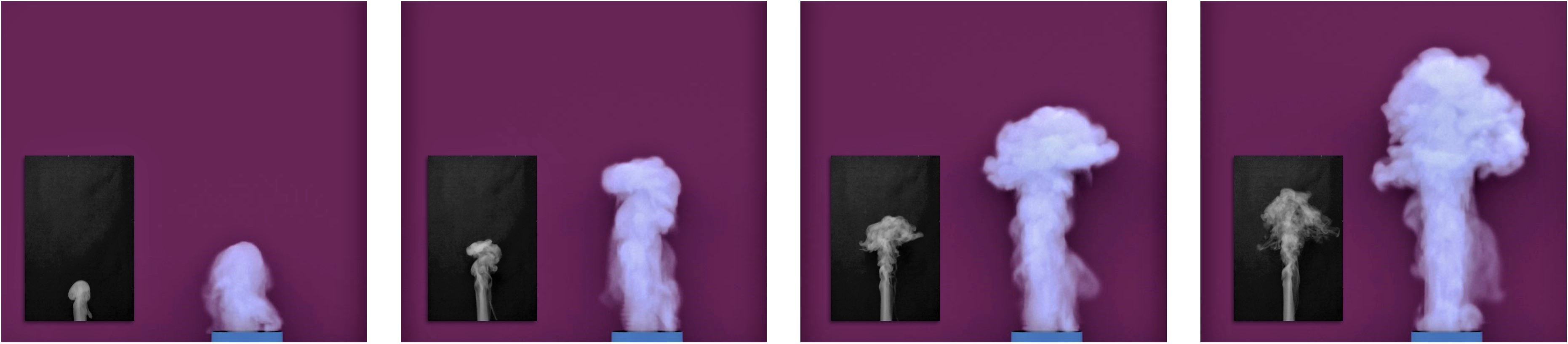}
  \caption{\nils{An example sequence of our ScalarFlow data set. Four frames of
    our data set are re-rendered as thick smoke. Insets of the corresponding 
	frames of one of the captured video streams, i.e. the real-world reference,
	are shown in the lower left corners.}}
\end{teaserfigure}

\maketitle

\input{1_introduction}
\input{2_relatedWork}
\input{3_method}
\input{6_capture}

\input{4_results}

\input{5_conclusion}

 \section*{Acknowledgments}
 This work was supported by the ERC Starting Grant {\em realFlow} (StG-2015-637014).
 We additionally thank Mengyu Chu for help with video editing.

\bibliographystyle{ACM-Reference-Format}
\bibliography{references}

\newpage
\clearpage

\appendix 

\small

\section{Additional Reconstructions}
\label{app:addRecon}

\noindent\begin{minipage}{\textwidth}
	\centering
	\includegraphics[width=.48\linewidth]{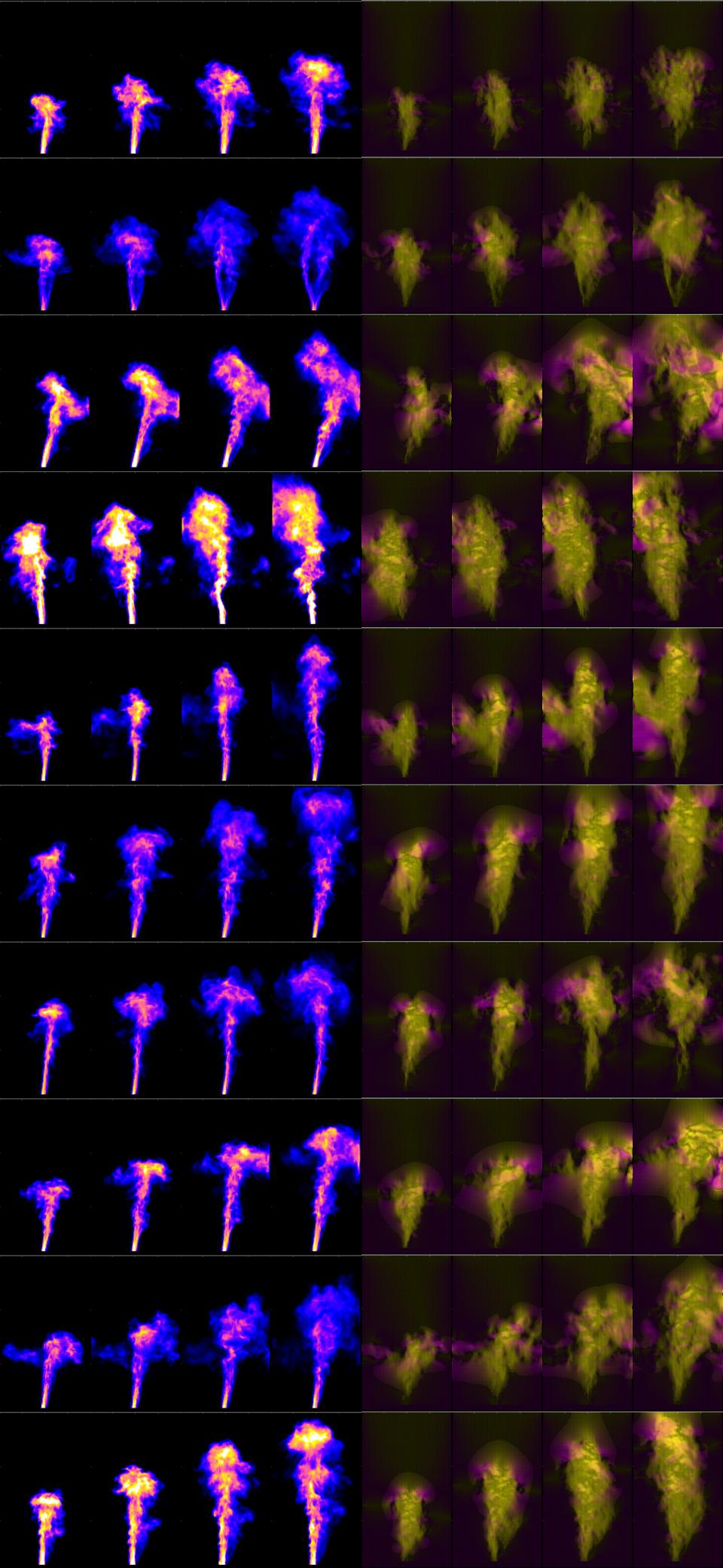}
	\includegraphics[width=.48\linewidth]{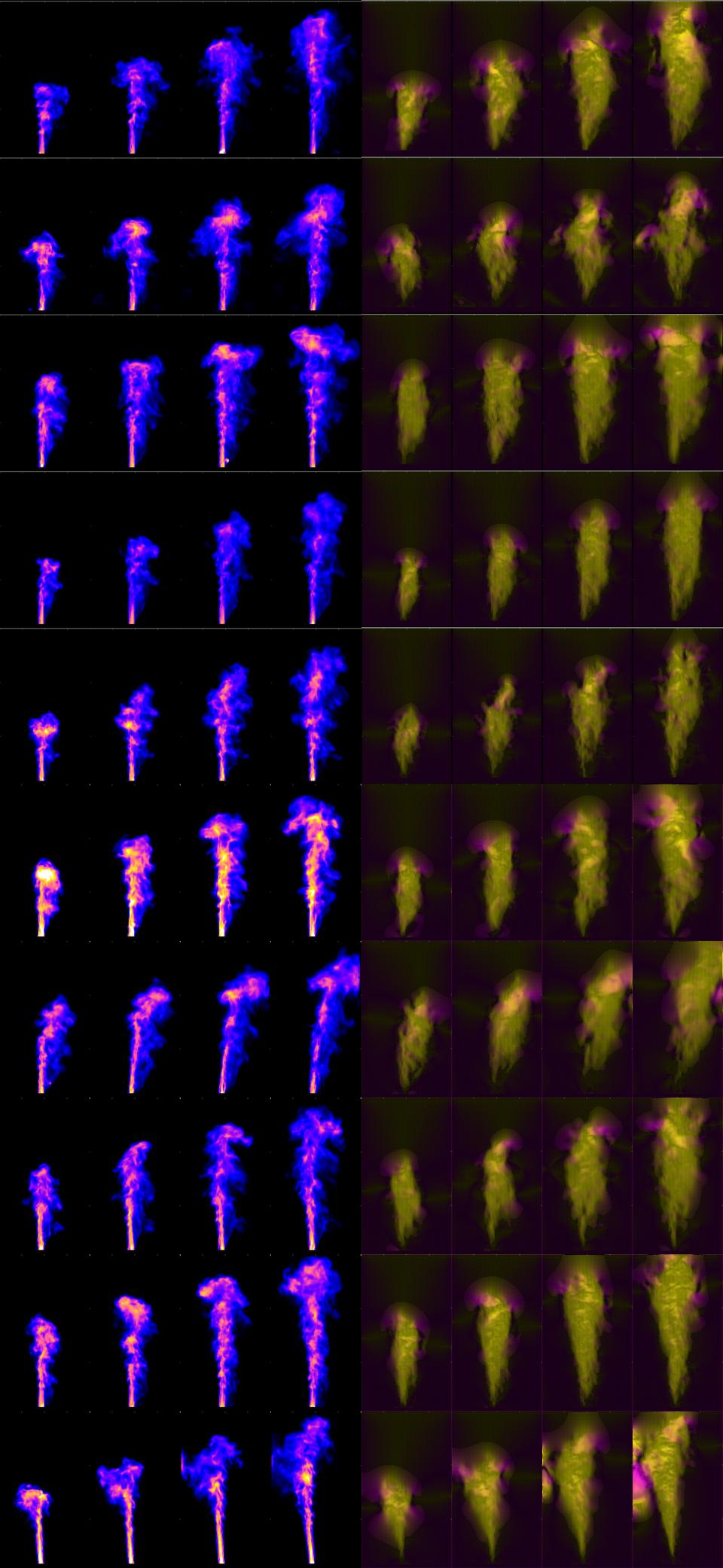}
	\captionof{figure}{A sub-set of captured clouds of the ScalarFlow data set visualized in
		terms of density and velocity. Each row on both columns shows four frames of
		a single capture.}
	\label{fig:collection}
\end{minipage}

\newpage
\clearpage

\section{Details of Reconstruction Algorithm}
\label{app:reconsAlg}

In this appendix, we outline our reconstruction algorithm with pseudocode where \myrefalg{alg:recProc} is the main routine. 
Based on an initial density estimate, it iteratively predicts the quantities' states and computes density inflow as well as a physics-constrained update to the velocities.
The tomography solve in \myrefalg{alg:tomo} is used for initialization and as a building block within the residual velocity calculation step of \myrefalg{alg:velUpd}. Lastly, \myrefalg{alg:velUpdMS}
details our recursive multi-scale version of the residual velocity computation, which makes use of 
\myrefalg{alg:velUpd}.

We add our smoothness and kinetic energy regularizers described in \myrefeq{eq:regularizers} to matrix $\textbf{A}$ for each linear system of equations (LSE) referred to in $\Call{solveLSE}{\textbf{A},\text{b}}$, 
i.e., for density and velocity in \\
$\Call{reconDen}{\i{t},\Phip{t}},$ $\Call{reconVel}{\up{t},\Phip{t},\i{t}}, $ and $\Call{estimInfl}{\i{t}, \Phy{t-1}, \u{t}}$.
Furthermore, in order to apply PD, a weighted identity matrix $\sigma \textbf{I}$ is added as well.

\begin{algorithm}[t]
	\caption{Full Reconstruction Procedure}
	\label{alg:recProc}
	\begin{algorithmic}[1]
		\Function{reconVelDen}{$\i{t}$}\Comment{reconstr. velocity and density}
		\State $\u{0}=0$; \Comment{init velocity}
		\State $\Phy{0} = \Call{reconDen}{\i{0},0}$; \Comment{init density, \myrefeq{eq:tomoFull}}
		\ForAll{t}
		\State \(\triangleright\) predict velocity
		\State $\up{t} = \Pi_{\text{DIV}}(\mathcal{A}(\u{t-1}, \u{t-1})) + \Call{addViscosity}{\u{t-1}}$;
		\State $\PhiIp{t} = \Call{estimInfl}{\i{t}, \Phy{t-1}, \up{t}}$; \Comment{estimate density inflow}
		\State $\Phip{t} = \mathcal{A}(\Phy{t-1}+\PhiIp{t}, \up{t})$; \Comment{predict density}
		\State \(\triangleright\) solve for residual velocity \myrefeq{eq:problem_statement}
		\State $\uu{t} = \Call{reconVelMS}{\up{t},\Phip{t},\Phy{t-1},\i{t}}$; 
		\State $\u{t}=\up{t}+\uu{t}$; \Comment{finalize velocity}
		\State $\PhiI{t} = \Call{estimInfl}{\i{t}, \Phy{t-1}, \u{t}}$; \Comment{estimate density inflow}
		\State $\Phy{t} = \mathcal{A}(\Phy{t-1}+\PhiI{t},\u{t})$; \Comment{finalize density}             
		\EndFor
		\EndFunction
	\end{algorithmic}
\end{algorithm}

\begin{algorithm}[t]
	\caption{Tomography, solve \myrefeq{eq:tomoFull}, \eqref{eq:tomoUpd}, or \eqref{eq:tomoCor}}
	\label{alg:tomo}
	\begin{algorithmic}[1]
		\Function{reconDen}{$\i{t},\Phi_{\text{IN}}^{t}$}\Comment{reconstr. density, $\Phi_{\text{IN}}^{t}$ can be 0}
			\State $\iu{t} = \i{t}-\PP\Phi_{\text{IN}}^{t}$; \Comment{calculate residual image}
			\For{$k\gets 1, 10$} \Comment{Prima-Dual iterations}
				\State \(\triangleright\) solve least-squares problem, e.g., with CGLS (\myrefalg{alg:regCGLS})
				\State $\x^k = \x^{k-1}+\sigma\y^{k-1}$
				\State $\quad -\sigma\Call{solveLSE}{\textbf{A}=\PPP, \text{b}=-\PP^T\iu{t}+\x^{k-1}+\sigma\y^{k-1}}$;
				\State $\z^k=\Pi_{\text{NonNeg}}(\Phi_{\text{IN}}^{t}+\z^{k-1}-\tau\x^k)$; \Comment{cut off neg. values}
				\State $\y^k=\z^k+\theta(\z^k-\z^{k-1})$; 
			\EndFor
			\State return $\z^k$;
		\EndFunction
	\end{algorithmic}
\end{algorithm}

\begin{algorithm}[t]
	\caption{Residual Velocity, solve \myrefeq{eq:problem_statement}}
	\label{alg:velUpd}
	\begin{algorithmic}[1]
		\State \(\triangleright\) calculate residual velocity on single scale
		\Function{reconVel}{$\up{t},\Phip{t},\i{t}$}
		\For{$k\gets 1, 10$} \Comment{Prima-Dual iterations}
		\State \(\triangleright\) shared part: advection equation, $\x^k=\xV^k\cup\xD^k$
		\State $\x^k = \x^{k-1}+\sigma\y^{k-1} -\sigma\Call{solveLSE}{\textbf{A}, \text{b}}$;
		\State \Comment{$ \textbf{A}=\begin{bmatrix}(\nabla\Phip{t})(\nabla\Phip{t})^T&(\nabla\Phip{t})\\(\nabla\Phip{t})^T&I\end{bmatrix}, \text{b}=\x^{k-1}+\sigma\y^{k-1}$}
		\State \(\triangleright\) velocity part: inflow and divergence-free on $\zV^{k}$ 
		\State $\uuI{t}=\zV^{k-1}-\tau\xV^k$; \Comment{create temporary velocity}
		\State $\uuI{t}|_{I} = c - \up{t}|_{I}$; \Comment{set velocity inflow}
		\State $\zV^k=\Pi_{\text{DIV}}(\uuI{t})$; \Comment{make div-free}
		\State \(\triangleright\) density part: tomo. and non-neg. (\myrefeq{eq:tomoCor}) on $\zD^{k}$ 
		\State $\zD^k=\zD^{k-1}-\tau\xD^k+\Call{reconDen}{\i{t},\Phip{t}+\zD^{k-1}-\tau\xD^k}$; 
		\State \(\triangleright\) shared variable update, $\y^k=\yV^k\cup\yD^k$
		\State $\y^k=\z^k+\theta(\z^k-\z^{k-1})$; 
		\EndFor
		\State return $\zV^k$;
		\EndFunction
	\end{algorithmic}
\end{algorithm}

\begin{algorithm}[t]
\caption{MS Residual Velocity, solve \myrefeq{eq:problem_statement}}
	\label{alg:velUpdMS}
	\begin{algorithmic}[1]
		\State \(\triangleright\) calculate residual velocity on multiple scales
		\Function{reconVelMS}{$\up{t},\Phip{t},\Phy{t-1},\i{t}$}
		\If{grid size large enough}
		\State $\uuC{t}=\Pi_{\text{DIV}}(\Call{up}{\Call{reconVelMS}{\Call{down}{\up{t}},\Call{down}{\Phip{t}},\i{t}}})$;
		\State $\up{t}_{\text{new}} = \up{t} + \uuC{t}$; \Comment{add coarse residual to prediction}
		\State $\PhiIpn{t} = \Call{estimInfl}{\i{t}, \Phy{t-1}, \up{t}_{\text{new}}}$; \Comment{den. inflow}	
		\State $\Phip{t}_{\text{new}} = \mathcal{A}(\Phy{t-1} + \PhiIpn{t}, \up{t}_{\text{new}})$; \Comment{adapt prediction}
		\State \(\triangleright\) sum up residual velocities on different scales
		\State $\uu{t} = \uuC{t} + \Call{reconVel}{\up{t}_{\text{new}},\Phip{t}_{\text{new}},\i{t}}$;
		\Else
		\State $\uu{t} = \Call{reconVel}{\up{t},\Phip{t},\i{t}}$; 
		\EndIf
		\State return $\uu{t}$;
		\EndFunction
	\end{algorithmic}
\end{algorithm}

\subsection{Visual Hull and Inflow Estimation}
\label{app:inflow}
For the density estimation (both tomography and inflow), we incorporate a visual hull to achieve higher accuracy and reduce computational complexity. 
Inflow source cells from $\Gamma_{RI}$ are set to zero and excluded from the solve 
in order to prevent density in cells that should be empty. 
We only mark them as outside if the contributing weight is above 1e-2,
while voxels are included in the system if their weight is above 1e-4. 
These thresholds were determined experimentally for our hardware setup and were used for all of our reconstructions.

Regarding the density discrepancy $d$ for the inflow solve, we could distribute the total missing
residual density from
$\Delta \Phy{t}_{\text{tar}}|_{\Gamma \setminus (\Gamma_R \cup \Gamma_I)}$ (positive
or negative) equally to each target cell, i.e.,
$$d=\frac{\sum\limits_{\Gamma \setminus (\Gamma_R \cup \Gamma_I)}
\Delta \Phy{t}_{\text{tar}}}{|\Gamma_R|}.$$ 
However, we instead choose to scale the value $d$ with
the value in $\Delta \Phy{t}_{\text{tar}}|_{\Gamma_R}$ such that a cell $e \in \Gamma_R$
with truly higher values gets a larger fraction of the missing densities assigned,
i.e.,
$$d_e=\frac{\sum\limits_{\Gamma \setminus (\Gamma_R \cup \Gamma_I)}
\Delta \Phy{t}_{\text{tar}}}
{\sum\limits_{\Gamma_R}(\Delta \Phy{t}_{\text{tar}}+\text{o})}(\Delta \Phy{t}_{\text{tar}}|_e+\text{o}),$$
where the offset o ensures that each $\Delta \Phy{t}_{\text{tar}}|_e\geq0$ for scaling.
We additionally require cells with negative values to contain less density than the positive ones. Accordingly,
the offset is 0 or the absolute value of the largest negative entry in
$\Delta \Phy{t}_{\text{tar}}|_{\Gamma_R}$, i.e., $\text{o}=|\text{min}(0,\Delta \Phy{t}_{\text{tar}}|_{\Gamma_R})|$.

\newpage
\onecolumn
\clearpage 
\section{Derivation of Optimization Problem}
\label{app:derivation}

\noindent In this section, we derive our optimization problem for the residual velocity step-by-step:
\begin{eqnarray}
	&\underset{\u{t},\Phy{t}}{\text{minimize}}
	&\norm{\frac{\partial \Phy{t}}{\partial t} + \nabla \Phy{t}\cdot \u{t}}^2  \nonumber \\
	&&\qquad \qquad \qquad \left| 
	\begin{array}{l}
		\Phy{t} = \mathcal{A}(\Phy{t-1}+\PhiI{t},\u{t}),\nonumber \\
		\u{t}=\up{t}+\uu{t}\\
	\end{array}
	\right.\\
	\equiv\; 
	&\underset{\uu{t},\PhiI{t}}{\text{minimize}}
	&\norm{\frac{\partial \mathcal{A}(\Phy{t-1}+\PhiI{t},\up{t}+\uu{t})}{\partial t} + \nabla \mathcal{A}(\Phy{t-1}+\PhiI{t},\up{t}+\uu{t})\cdot (\up{t}+\uu{t})}^2  \nonumber \\
	&&\qquad \qquad \qquad  \left| 
	\begin{array}{l}
	\text{linearize advection}:\nonumber \\
	\mathcal{A}(\mathcal{A}(\Phy{t-1}+\PhiI{t},\up{t}),\uu{t})\approx\mathcal{A}(\Phy{t-1}+\PhiI{t},\up{t}+\uu{t})\nonumber \\
	\end{array}
	\right.\nonumber \\
	\equiv\; 
	&\underset{\uu{t},\PhiI{t}}{\text{minimize}}
	&\norm{\frac{\partial \mathcal{A}(\mathcal{A}(\Phy{t-1}+\PhiI{t},\up{t}),\uu{t})}{\partial t} + \nabla \mathcal{A}(\mathcal{A}(\Phy{t-1}+\PhiI{t},\up{t}),\uu{t})\cdot (\up{t}+\uu{t})}^2  \nonumber \\
	&&\qquad \qquad \qquad \left| 
	\begin{array}{l}
	\PhiIp{t}=\PhiI{t},\nonumber \\
	\nabla \Phip{t}\approx \nabla \Phy{t}\\
	\end{array}
	\right.\nonumber \\
	\equiv\; 
	&\underset{\uu{t},\Phy{t}}{\text{minimize}}
	&\norm{\frac{\partial \mathcal{A}(\mathcal{A}(\Phy{t-1}+\PhiIp{t},\up{t}),\uu{t})}{\partial t} + \nabla \mathcal{A}(\Phy{t-1}+\PhiIp{t},\up{t})\cdot (\up{t}+\uu{t})}^2  \nonumber \\
	&&\qquad \qquad \qquad \left| 
	\begin{array}{l}
	\Phip{t} = \mathcal{A}(\Phy{t-1}+\PhiIp{t}, \up{t}),\nonumber \\
	\text{remove constant terms}\\
	\end{array}
	\right.\nonumber \\
	\equiv\; 
	&\underset{\uu{t},\Phy{t}}{\text{minimize}}
	&\norm{\frac{\partial \mathcal{A}(\Phip{t},\uu{t})}{\partial t} + \nabla \Phip{t} \cdot \uu{t}}^2 \nonumber \\
	&&\qquad \qquad \qquad \left| 
	\begin{array}{l}
	\Phiu{t}=\frac{\partial \mathcal{A}(\Phip{t},\uu{t})}{\partial t} = \frac{\Phip{t}+\Phiu{t}-\,\Phip{t}}{\Delta t}, \Delta t=1 \nonumber \\
	\end{array}
	\right.\nonumber \\
	\equiv\; 
	&\underset{\uu{t},\Phiu{t}}{\text{minimize}}\Aboxed{&\norm{\Phiu{t} + \nabla \Phip{t} \cdot \uu{t}}^2 } \label{eq:opti0}\\
%%%%%%%%%%%%%%%%%%%%%%%%%%%%%%%%%%%
&\text{subject to }
&\u{t}|_{I} = c \qquad \qquad \qquad \! \left| \; \u{t}=\up{t}+\uu{t}, \text{ where \textit{c} is inflow speed} \right. \nonumber \\
\equiv\; 
&&\up{t}|_{I} + \uu{t}|_{I}= c \qquad \left| \; \up{t} \text{ is div-free} \right. \nonumber \\
\equiv\; 
&\Aboxed{&\uu{t}|_{I} = c - \up{t}|_{I},} \label{eq:opti1} \\
%%%%%%%%%%%%%%%%%%%%%%%%%%%%%%%%%%%
&&\nabla \cdot \u{t} = 0 \qquad \qquad \qquad \! \left| \; \u{t}=\up{t}+\uu{t} \right. \nonumber \\
\equiv\; 
&&\nabla \cdot (\up{t}+\uu{t}) = 0 \qquad \left| \; \up{t} \text{ is div-free} \right. \nonumber \\
\equiv\; 
&\Aboxed{&\nabla \cdot \uu{t} = 0,} \label{eq:opti2}\\
%%%%%%%%%%%%%%%%%%%%%%%%%%%%%%%%%%%
&&\PP \Phy{t} - \i{t} = 0 \qquad \left| \; \Phy{t} = \mathcal{A}(\Phy{t-1}+\PhiI{t},\u{t}) \right. \nonumber \\
\equiv\; 
&&\PP \mathcal{A}(\Phy{t-1}+\PhiI{t},\u{t}) - \i{t} = 0 \qquad \left| \;  \u{t}=\up{t}+\uu{t}, \PhiIp{t}\approx\PhiI{t}, \text{ lin. adv.} \right. \nonumber \\
\equiv\; 
&&\PP \mathcal{A}(\mathcal{A}(\Phy{t-1}+\PhiIp{t},\up{t}),\uu{t}) - \i{t} = 0 \qquad \left| \; \Phip{t} = \mathcal{A}(\Phy{t-1}+\PhiIp{t}, \up{t}) \right. \nonumber \\
\equiv\; 
&&\PP \mathcal{A}(\Phip{t},\uu{t}) - \i{t} = 0 \qquad \left| \; \mathcal{A}(\Phip{t},\uu{t}) = \Phip{t}+\Phiu{t} \Rightarrow \Phy{t}\approx\Phip{t}+\Phiu{t} \right. \nonumber \\
\equiv\; 
&&\PP \Phip{t} + \PP \Phiu{t} - \i{t} = 0,  \qquad \left| \; \ip{t}=\PP \Phip{t} \right. \nonumber \\
\equiv\; 
&&\PP \Phiu{t} - (\i{t}-\ip{t}) = 0,  \qquad \left| \; \iu{t}=\i{t}-\ip{t}\right. \nonumber \\
\equiv\; 
&\Aboxed{&\PP \Phiu{t} - \iu{t} = 0,} \label{eq:opti3}\\
%%%%%%%%%%%%%%%%%%%%%%%%%%%%%%%%%%%
&&\Phy{t} \geq 0 \qquad \qquad \qquad \left| \; \Phy{t} \approx \Phip{t}+\Phiu{t} \right. \nonumber \\
\equiv\;  
&&\Phip{t}+\Phiu{t} \geq 0 \nonumber \\
\equiv\; 
&\Aboxed{&\Phiu{t} \geq -\Phip{t}}\label{eq:opti4}.
\end{eqnarray}

\clearpage

\noindent The problem formulations for calculating density, residual density, or density correction are
\begin{align}
&\underset{\Phy{t}}{\text{minimize}} \norm{P\Phy{t}-\i{t}}^2, \text{subject to } \Phy{t}\geq0  \label{eq:tomoFull},  \\ 
&\underset{\Phiu{t}}{\text{minimize}} \norm{\PP\Phiu{t}-(\i{t}-\PP\Phip{t})}^2,\text{subject to } \Phiu{t}\geq-\Phip{t} \label{eq:tomoUpd},
\\ 
\Aboxed{&\underset{\Phic{t}}{\text{minimize}} \norm{\PP\Phic{t}-(\i{t}-\ip{t}-\PP\Phiu{t})}^2, \text{subject to } \Phic{t}\geq-\Phiu{t}-\Phip{t}} \label{eq:tomoCor}.
\end{align}

\vspace{0.5cm}

\noindent The problem formulation for the density inflow estimation is described as following :
\begin{align}
\sum\limits_{e \in \Gamma \setminus \Gamma_I} \Phy{t} &\overset{!}{=} \sum\limits_{e \in \Gamma \setminus \Gamma_I} \Phy{t}_{\text{tar}} \nonumber\\
\sum\limits_{e \in \Gamma \setminus \Gamma_I} \mathcal{A}(\Phy{t-1}+\PhiI{t},\u{t}) & \overset{!}{=} \sum\limits_{e \in \Gamma \setminus \Gamma_I} \Phy{t}_{\text{tar}} \qquad \qquad \qquad \qquad |\triangleright \text{ assume linear advection } \nonumber\\
\sum\limits_{e \in \Gamma \setminus \Gamma_I} \mathcal{A}(\Phy{t-1},\u{t}) + \sum\limits_{e \in \Gamma \setminus \Gamma_I} \mathcal{A}(\PhiI{t},\u{t}) & \overset{!}{=} \sum\limits_{e \in \Gamma \setminus \Gamma_I} \Phy{t}_{\text{tar}} \qquad \qquad \qquad \qquad|\triangleright \text{ move to other side } \nonumber\\
\sum\limits_{e \in \Gamma \setminus \Gamma_I} \mathcal{A}(\PhiI{t},\u{t})& \overset{!}{=} \sum\limits_{e \in \Gamma \setminus \Gamma_I} \big(\Phy{t}_{\text{tar}} - \mathcal{A}(\Phy{t-1},\u{t}) \big) \:\:\:\:\:\: |\triangleright \text{ define }\Delta \Phy{t}_{\text{tar}}=\Phy{t}_{\text{tar}} - \mathcal{A}(\Phy{t-1},\u{t})\nonumber\\
\sum\limits_{e \in \Gamma_R} \mathcal{A}(\PhiI{t},\u{t})& \overset{!}{=} \sum\limits_{e \in \Gamma \setminus \Gamma_I} \Delta \Phy{t}_{\text{tar}}.
\end{align}
With $\mathcal{A}(\PhiI{t},\u{t})$, we only influence $\Gamma_R$, but not the full visible domain $\Gamma \setminus \Gamma_I$. 
Therefore, each cell $e \in \Gamma_R$ is the sum of the residual target density plus an offset $d$ that accounts for further discrepancies in $\Gamma \setminus \Gamma_R \setminus \Gamma_I $, which is used as right-hand side b in \myrefalg{alg:estimInflow}. 
\begin{align}
\Aboxed{\mathcal{A}(\PhiI{t},\u{t})|_e = \text{max }\left(\Delta \Phy{t}_{\text{tar}}|_e + d, -\mathcal{A}(\Phy{t-1},\u{t})|_e \right)}, \label{eq:inflowDerivation}
\end{align}

\end{document}

%% file: 0_abstract.tex
In this paper, we present {\em ScalarFlow}, a first large-scale data set of
reconstructions of real-world smoke plumes. In addition, we propose a framework
for accurate physics-based reconstructions from a small number of video streams.
Central components of our framework are a novel estimation of unseen inflow
regions and an efficient optimization scheme
constrained by a simulation to capture real-world fluids.
Our data set includes a large
number of complex natural buoyancy-driven flows. The flows transition to
turbulence and contain observable scalar transport processes. As such, the
ScalarFlow data set is tailored towards computer graphics, vision, and learning
applications. The published data set contains volumetric reconstructions of
velocity and density as well as the corresponding input image sequences with calibration
data, code, and instructions how to reproduce the commodity hardware capture
setup. We further demonstrate one of the many potential applications: a
first perceptual evaluation study, which reveals that the complexity of the
reconstructed flows would require 
large simulation resolutions for regular solvers in order to recreate at least
parts of the natural complexity contained in the captured data.

%%% Local Variables:
%%% mode: latex
%%% TeX-master: "paper"
%%% End:

%% file: 1_introduction.tex
\section{Introduction}

\nt{Despite the long-standing success of physical simulations as tools for visual effects (VFX)
production, there is a notable lack of benchmark cases and data sets for
evaluating the simulated results. 
While other fields of research, such as computational
photography and geometry processing, can rely on large image and model databases
for evaluation as well as machine learning applications, simulations for visual effects 
typically are only evaluated in terms of visuals by a small number of
experts. This is partially caused by the inherent difficulties 
of capturing real-world counterparts for the simulated phenomena. Typically,
we are dealing with volumetric effects, such as clouds of smoke, liquids, or
deformable bodies, that inherently require a full acquisition of the three-dimensional (3D) volume 
and its motion. Thus, it is crucial to obtain a \me{3D} description of
the configuration of a material and its velocity field in order to compare and 
align the corresponding simulated quantities.}

\begin{figure*}[t!]
	\center
	\includegraphics[width=1.0\textwidth]{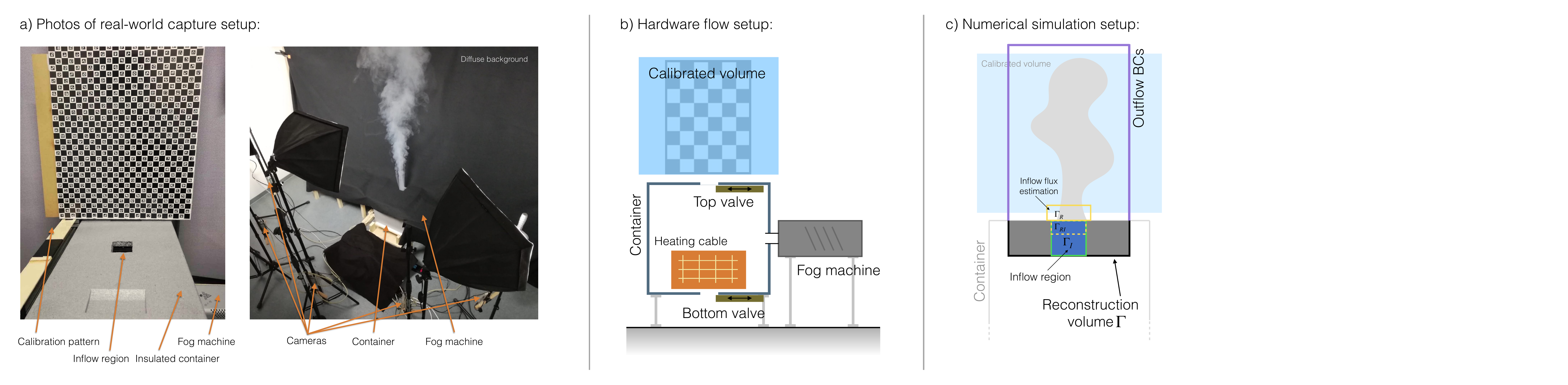}
	\caption{\nt{An overview of our capture setup: the container for smoke, calibration plate and camera setup (a), 
		an illustration of the core components (b), 
		and the corresponding numerical simulation and reconstruction domain (c).}}
	\label{fig:setup2}
\end{figure*}

In addition, the advent of data-driven methods, deep learning techniques in
particular, has demonstrated the possibilities that arise from the availability
of data sets. Most famously, the ImageNet data set \cite{deng2009imagenet} has
been used in thousands of studies and led to huge advances for image classification. 
By now, many other data sets exist ranging from videos
for action recognition \cite{abu2016youtube} to 3D scenes
for geometry reconstruction \cite{knapitsch2017tanks}. These data sets
illustrate the potential for invigorating research, encouraging reproducible
evaluations, and reducing the barrier of entry for newcomers. Overall, the
availability of 
% carefully prepared, 
reliable data sets with sufficient
complexity has led to significant progress of research in the corresponding
fields. While first studies have successfully established connections between learning
methods with physics-based simulations
\cite{Ladicky2015,chu2017cnnpatch,ma2018rbrl}, all of the studies so far purely
rely on case-specific and synthetic (i.e., purely simulated) data.

Our goal with this work is to provide a first large-scale data set of real smoke
clouds. The data was captured with a sparse, physics-based, multi-view tomography
and contains volumetric reconstructions (i.e., velocity and density fields),
video data, and camera calibrations. In addition, source code for capture and
reconstruction will be published.

The primary contribution of our work is the creation of a carefully designed
data set with a large number of accurate reconstructions of real-world scalar
transport phenomena. Beyond this primary goal, 
our work contains several technical contributions:
we propose the design of a low-cost hardware setup and an efficient optimization 
scheme for tomographic reconstructions. \me{Our approach is able to recreate
real-world smoke behavior even on modest resolutions by guiding a simulator
along real-world footage and by employing an advanced}
method for estimating inflow boundary conditions.
We also present, as a very first model application, a perceptual user study of different
simulation methods and resolutions for buoyant smoke clouds based on our data
set. 
{Interestingly, these studies reveal that our reconstructions 
contain detail and natural dynamics that 
require forward simulations with very high resolutions in order to be matched.}

%%% Local Variables:
%%% TeX-master: "paper"
%%% End:

%% file: 2_relatedWork.tex
\section{Related Work}

{\em Scientific Data Sets} have a long history in research, and a popular
example from the computational fluid dynamics (CFD) community is the Johns
Hopkins Turbulence Database. It contains eight very finely resolved simulations
of different turbulent flows \cite{li2008public}. Despite being widely used for
turbulence modeling, the data sets in this database focus on non-visual flows,
i.e., without observable quantities, and contain a single data set for each of
the eight setups. Hence, the data is not readily usable for graphics, vision, or
deep learning applications. Other CFD and fluid flow databases, such as the THT
Lab \cite{myong1990new}, KTH Flow \cite{schlatter2010dns}, and FDY DNS
\cite{avsarkisov2014fdy} databases, share these characteristics: they offer
single simulated data sets with varying resolutions.

With our database, we target scalar transport processes in the form of buoyant
plumes of hot smoke, which represent a large class of fluid problems in computer
animation \cite{fedkiw:2001:VSO,sato2018example}. At the same time, scalar transport
phenomena are important for engineering \cite{moin1991dynamic,yazdani2018hidden}
and medical applications \cite{morris2016heartCfd} \nils{because the transported 
quantities yield important information about the flow motion.
We specifically aim for providing many varied data sets of a single phenomenon, e.g., 
to support the construction of reduced models.}

\nt{{\em Fluid Simulations} are important and established components in \nt{numerous
fields \cite{harlow1965,cummins2018separated}. 
In graphics, a popular Eulerian, i.e. grid-based, solver for flow simulations is the 
{\em Stable Fluids} algorithm \cite{Stam1999} and 
various extensions are available, e.g., to retain kinetic energy
\cite{fedkiw2001visual,selle2005vortex}, to speed up the pressure projection component \cite{McAdams:2010:PMP,setaluri2014spgrid},
and to improve the accuracy of the advection step \cite{Kim05FlowFixer,Selle:2008:USM,zehnder2018}.}
For many VFX applications, guiding and control \cite{ShiTamingLiquids,nielsen2009guiding,pan2017efficient}
are highly important.
%\
In addition, the material point method %, on the other hand, 
has become a popular alternative for complex, fluid-like materials
\cite{Stomakhin:2013:MPM,jiang2016mpmCourse,tampubolon2017multi}, while the
class of smoothed particle hydrodynamics methods represent purely Lagrangian
variants \cite{IhmsenOSKT14a}. 
Additionally, vortex filaments \cite{angelidis2006controllable,Weissmann:2010:FSV:1778765.1778852}
are a popular Lagrangian representation for volumetric flows.}
We will focus on Eulerian solvers as they are
widely used for single-phase smoke simulations. A thorough overview can be found
in corresponding text books \cite{bridson2015}.
Note that while graphics publications typically refer to dense, passively
advected tracers as {\em smoke}, our hardware setup diffuses water and, as such,
produces {\em fog}. However, this is purely a difference in terminology; the
underlying physics are equivalent. Thus, our setup is representative for the
commonly used hot smoke simulations \cite{fedkiw2001visual}, 
and we will refer to the tracers as smoke from now on.

{\em Flow Capture and Tomography:} Fluid flows are inherently difficult to capture. While
traditional experiments often perform only very localized measurements of
quantities like pressure and velocity \cite{kavandi1990luminescent}, \nt{methods
such as {\em particle image velocimetry} (PIV) \cite{Elsinga:06} can be used to
recover volumetric information about flow motions by injecting particles. 
While specialized methods were able to recover 3D
flows over time \cite{Xiong:2017}, PIV cannot be employed in graphics settings
as substances in the fluid such as smoke would immediately obscure the
particles and the process is typically restricted to relatively small volumes.}

\nt{While laser scanning setups \cite{Hawkins:timeVary},
sheet-based reconstructions \cite{hasinoff2007photo},
and Schlieren-based capture algorithms \cite{atcheson2008time,Atcheson:2008:OEF}
have been proposed,
tomographic reconstructions are a more widely used alternative. In this
case, external observations, typically in the form of two-dimensional \me{(2D)} images,
are used to recover a 3D density. Beyond the medical field, such methods were
successfully used to, e.g., reconstruct flames \cite{Ihrke:tomoFlames,Ihrke:adaptive} 
and nebulae \cite{wenger2013fast}. 
Higher-level optimization involving tomographies were also proposed for the detailed capture 
of slowly deforming objects \cite{zang2018space,zang2018super}.}
We likewise make use of tomographic reconstructions from a sparse set of views, 
which are combined with physical constraints to obtain a realistic flow field. Algorithms
employing physical simulations (or parts thereof) were successfully used in
previous work to capture liquids \cite{wang2009physically}, divergence-free
motions \cite{Gregson:2014}, and smoke volumes via appearance transfer
\cite{Okabe:2015}.
Similar in 
spirit to our work, a physics-based single-view reconstruction 
was proposed \cite{eckert18}, which, however, lacks
our procedure for inflow estimation and does not yield sufficiently reliable
reconstructions due to being restricted to a single viewpoint. 

{\em Applications Areas:} There are numerous potential areas of application for
\nt{volumetric flow data sets in the areas of computer vision, graphics, 
and machine learning alone. 
To leverage machine learning in the context of simulations,
researchers have, e.g., used machine learning to drive particle-based simulations \cite{Ladicky2015},}
replaced the traditional pressure solve with pre-trained models
\cite{tompson2016accelerating}, and augmented simulated data with learned descriptors
\cite{chu2017cnnpatch}. Others have focused on learning controllers for rigid
body interactions \cite{ma2018rbrl} or aimed for temporal coherence with
adversarial training \cite{xie2018tempogan}. This is a nascent field with
growing interest, where our data set can provide a connection of simulations
with the real world.
Although this is a promising direction, we will focus 
on evaluating the results of established simulation 
techniques in comparison to real-world flows as a first
demonstration of the usefulness of a volumetric flow data set.
Perceptual evaluations have been used to evaluate rendering algorithms \cite{cater2002},
tone mapping \cite{masia2009}, and animations of human characters
\cite{hoyet2013}. Recently, different methods for liquid simulations were also
evaluated in regard to their visual realism \cite{um2017perceptual}. However, to the best of our knowledge, no
perceptual evaluations of the more fundamental algorithms for single-phase flows exist so
far. 

%%% Local Variables:
%%% TeX-master: "paper"
%%% End:

%% file: 3_method.tex
\section{Method}\label{sec:method}

\begin{figure}[tb]
	\centering
	\captionsetup[subfigure]{aboveskip=7pt,belowskip=7pt}
	\subcaptionbox{Typical steps of a fluid 
		simulator.\label{fig:simOverview}}{\includegraphics[width=\linewidth]{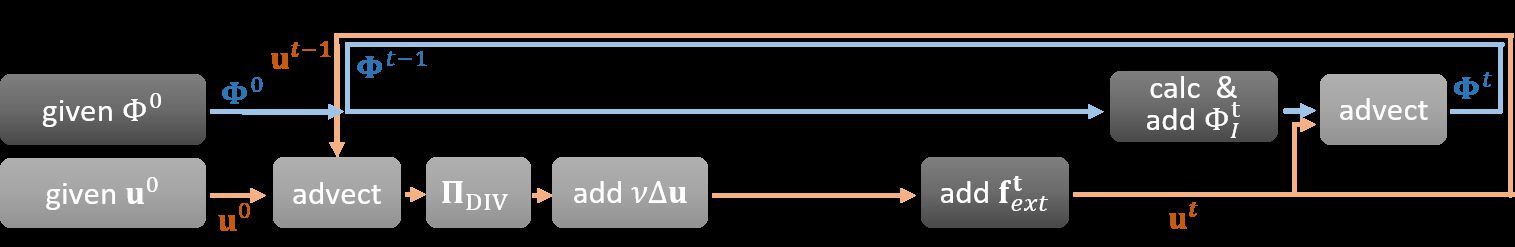}}\\
	\subcaptionbox{Overview of our reconstruction 
		method.\label{fig:reconOverview}}{\includegraphics[width=\linewidth]{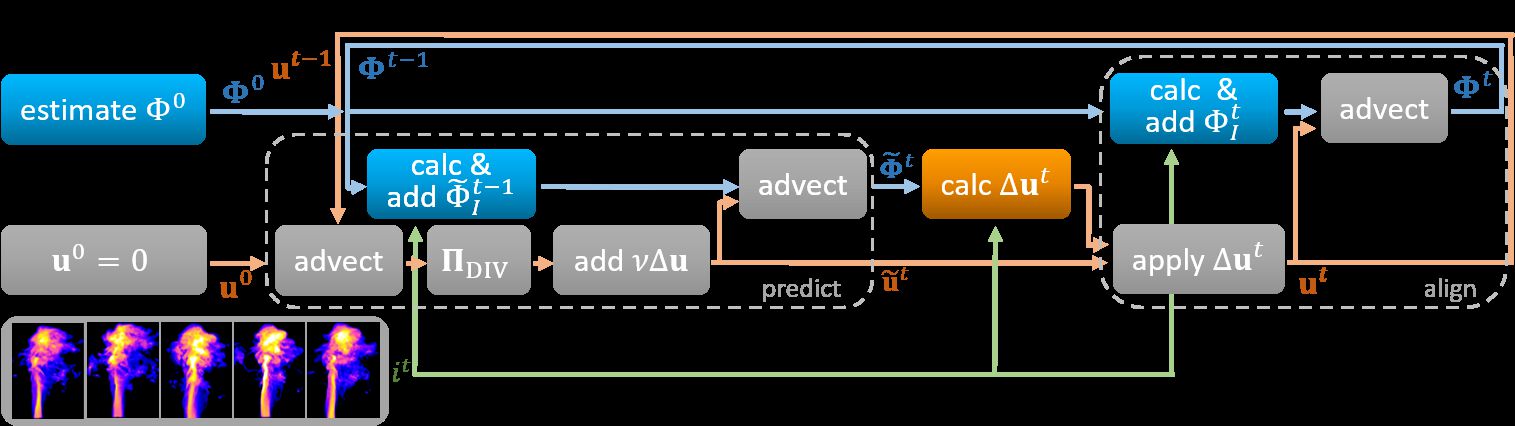}}
	\caption{\nt{Visual summary of a typical forward fluid simulation and its inverse 
			problem of recovering a simulation that matches a set of given target images. 
			%\me{ leading a simulation towards given target images.   % leading/guiding sounds artificial?
			While we can re-use many steps from a regular simulator for the latter, i.e. advection, 
			pressure projection, and viscosity, we need to estimate the unknown initial quantities, 
			especially the 3D density, estimate a valid density inflow in compliance with the target images, and account for any unknown force.}}
	\label{fig:cmpSimRecon}
\end{figure}

In the following, we will outline the reconstruction algorithm that was used to create the
\acro{} data set. 
A preview of our physical setup for capturing real-world smoke plumes is shown 
in \myreffig{fig:setup2} and will be explained in more detail in \myrefsec{sec:hardware}.
Our goal is to jointly reconstruct smoke density $\PhyNoT$ and velocity $\uNoT$ 
from a small number of image sequences $\iNoT$. \nils{This represents an inverse 
problem \me{as} our goal is to find a solution of a numerical simulation that
matches a set of observations. 
The widely established physical model for fluids are the incompressible Navier-Stokes equations,} written as
\begin{equation}
\begin{aligned}
\frac{\partial \uNoT}{\partial t} + \uNoT \cdot \nabla \uNoT &= - \nabla p + \nu\triangle\uNoT + \fNoT \\
  \nabla \cdot \uNoT &= 0,
\end{aligned}
\label{eq:Euler}
\end{equation}
where $\uNoT$ is velocity, $p$ is pressure, $\fNoT$ are external body forces, and $\nu$ is kinematic viscosity. 
Typical simulators solve those equations for each time step $t$ via operator splitting: the velocity is advected forward, projected onto the divergence-free space via the so-called pressure solve, viscosity and external forces are added, and optionally, inflow quantities are added. 
A schematic overview can be found in \myreffig{fig:simOverview}. 
Note that common simulators often exhibit a limit on resolvable features and, as such, sub-grid effects such as turbulent mixing are not directly captured. 

It is inherently difficult to re-simulate real-world fluid phenomena as initial density distribution $\Phy{0}$, density inflow $\PhiI{t}$, initial velocity $\u{0}$, velocity inflow, and ambient air motions are unknown.
External forces $\fNoT$ such as buoyancy are likewise typically difficult to model.
Furthermore, simulators introduce numerical errors, which need to be counteracted, particularly for moderate domain resolutions, in order to reproduce the turbulent motion found in real-world smoke plumes.
\nils{In this context, our approach shares similarities with methods for flow guiding \cite{nielsen2009guiding,Pan:2013}.
To match the real-world capture, we account for all unknown changes of velocities, i.e., forces, by}
calculating a residual velocity $\uu{t}$ and a density inflow $\PhiI{t}$.

Our full reconstruction algorithm is visualized in \myreffig{fig:reconOverview} and described in \myrefapp{app:reconsAlg} in \myrefalg{alg:recProc} with pseudocode.
We also provide algorithms in pseudocode for all of the following steps.
In our reconstruction algorithm, we first estimate the initial density volume $\Phy{0}$ with a single pass of regular
tomography and assume the fluid is initially at rest, i.e., $\u{0}=0$. 
Then, in order to ensure temporal coherence, we predict velocity $\up{t}$ by advecting the previous
velocity $\u{t-1}$ forward, making it divergence-free, and applying viscosity. 
To obtain a prediction for density, we calculate and add to $\Phy{t-1}$ an inflow source 
$\PhiIp{t}$ before advecting $\Phy{t-1}$ with $\up{t}$. Details of this step will be given in \myrefsec{sec:inflow}.
The predicted density $\Phip{t}$ is an intermediate variable that is only required 
to calculate the residual velocity, as shown in \myreffig{fig:reconOverview}. 
Our reconstruction method proceeds by accounting 
for the residual velocity that is necessary to match the motion and shape from the input image 
sequences by computing $\uu{t}$.
In an alignment step, $\uu{t}$ and $\up{t}$ are accumulated to yield the 
final velocity $\u{t}$. 
For computing the final density field $\Phy{t}$, it is crucial to estimate the 
correct amount of inflow density $\PhiI{t}$, which is added to $\Phy{t-1}$ before it is advected 
with $\u{t}$ to finally obtain $\Phy{t}$.
It is important to emphasize that, similar to a regular forward simulation, 
the smoke density is solely changed via advection. % and by adding a source in the invisible inflow region. 
We never modify densities outside the inflow region directly.
We denote our reconstruction domain with $\Gamma$, while the inflow region is denoted with $\Gamma_I$, which is visualized in \myreffig{fig:setup2}c.
%We use 
Outflow boundary conditions are set at the domain sides. Here velocity is allowed to move freely and densities in this 
region are removed.

\subsection{Residual Velocity Estimation}\label{sec:motionrecons}

\me{We calculate a residual velocity $\uu{t}$ that moves the predicted density $\Phip{t}$ such that
the input images $\i{t}$ are matched. 
We denote the density change induced by $\uu{t}$ with $\Phiu{t}$. \nils{This change in density
is an important intermediate variable required to enforce constraints on the density, \me{ which then constrains the residual velocity through our combined optimization scheme.}} 
As formulated in \myrefeq{eq:problem_statement}, we use the following physical 
assumptions for our reconstruction approach:
residual velocity  $\uu{t}$ and density $\Phiu{t}$ must comply with the transport equations; 
the total velocity inflow $\up{t}|_{I}+\uu{t}|_{I}$ is equal to a constant $c$;
the residual velocity field is incompressible;
projecting the residual density back to each camera plane should be equal 
to the difference between input images $\i{t}$ and the \nils{current, projected density prediction $\ip{t}$. 
Finally, the sum of residual and predicted density must be non-negative. This yields
the following minimization problem:}
\begin{equation}
\begin{aligned}
  & \underset{\Phiu{t},\uu{t}}{\text{minimize}}
  & & g(\uu{t},\Phiu{t}) = \norm{\Phiu{t}+\nabla\Phip{t}\cdot\uu{t}}^2\\
  & \text{subject to} & & \PP\:\Phiu{t}=\i{t}-\ip{t},
  \; \Phiu{t}+\Phip{t}\geq 0,\\ 
  &&&\uu{t}|_{I} = c - \up{t}|_{I}, \nabla \cdot \uu{t}=0,
\end{aligned}
\label{eq:problem_statement}
\end{equation}
where $\PP$ is the matrix projecting 3D density back to each 2D image, 
the finite difference for $\frac{\partial\Phip{t}}{\partial t}$ is denoted with $\Phiu{t}$, and
$c$ is a constant approximating the average upwards speed of the \nils{observed flow.} %input phenomenon.
A detailed derivation of \myrefeq{eq:problem_statement} can be found in \myrefapp{app:derivation}. % todo , add label
%the \highlight{appendix} in \myrefeq{eq:opti0} - (\ref{eq:opti4}).

% how solve
Following common practice \cite{parikh2013proximal}, we make our problem formulation convex by adding smoothness 
and kinetic energy regularizers for both density and velocity, namely 
\begin{equation}
\begin{aligned}
&E_{\text{smooth}}(\Phiu{t})&=&\half \norm{\nabla (\Phiu{t})}^2,  
&&E_{\text{smooth}}(\uu{t})&=&\half \norm{\nabla \uu{t}}^2,  \\
&E_{\text{kin}}(\Phiu{t})&=&\half \norm{\Phiu{t}}^2, 
&&E_{\text{kin}}(\uu{t})&=&\half \norm{\uu{t}}^2. 
\end{aligned}
\label{eq:regularizers}
\end{equation}
The kinetic and smoothness regularizers are realized by adding a weighted diagonal and a Laplacian matrix to the system matrix, respectively. 
The Laplacian operator is discretized with central finite differences, yielding the standard 7-point stencil.
%matrix contains $7$ entries in 3D for the central element and its neighbors in all directions (6 and -1).
We also regularize the density inflow.
The weights for both regularizers depend on the actual values of density and velocity. 
In our case, we used ((1e-1, 5e-4), (6e-1, 5e-2), (5e-3, 1e-2)) for smoothness and kinetic regularizers for velocity, density, and inflow density, respectively.

\nils{We compute solutions for the residual velocity's \me{objective function} %subproblem
%objective function for tomography 
$g(\Phiu{t},\uu{t})$ with least-squares and 
a standard Conjugate Gradient (CG) solver.}
As the shared objective function $g$ is convex and we have separate orthogonal 
projections for both unknowns $\Phiu{t}$ and $\uu{t}$, we use 
a fast primal-dual algorithm (PD) \cite{ChambollePD} % we can bring back later :) ... ,Inglis:2017,eckert18}. 
to solve the joint optimization problem.
The divergence-free constraint on the velocity is an orthogonal projection onto 
the space of divergence-free vector fields as shown in \cite{Gregson:2014}. 
This is enforced via a regular pressure solver. 
Before enforcing incompressibility, we set our inflow residual velocity to the difference between the prescribed constant $c$ and the predicted velocity $\up{t}$.
In order to fulfill both constraints on the residual density $\Phiu{t}$, we 
project it onto the space of densities where the input images are matched and the total density is non-negative. 
This can be realized by computing a density correction $\Phic{t}$, which is added 
to $\Phiu{t}$, as will be explained in \myrefsec{sec:tomo}.
In line with optical flow (OF) methods \cite{meinhardt2013horn}, we solve for 
the residual velocity on multiple scales, i.e. employ a hierarchical approach as outlined in \myrefalg{alg:velUpdMS}.}

% TODO discretization? staggered OF

While previous work has attempted single-view reconstructions with a similar
algorithm, we leverage the described physics-con\-strained optimization scheme for
accurate multi-view tomographies. Using multiple views has the advantage that fewer
unknown motions exist in the volume; thus, we do not need to employ the {\em depth-regularization} from previous work \cite{eckert18}. 
Next, we will explain our novel inflow estimation step and our efficient tomography
scheme, which is important for obtaining feasible run times for the reconstruction.

\subsection{Regularized Tomography} 
\label{sec:tomo}
%Alg. 1: why is this algorithm guaranteed to yield a solution that satisfies the constraint?

In order to satisfy both the image-matching and non-negativity constraints for
the residual density $\Phiu{t}$, we add a correction $\Phic{t}$ by solving a
second minimization problem:
\begin{equation}
\begin{aligned}
& \underset{\Phic{t}}{\text{minimize}}
& & h(\Phic{t}) = \norm{\PP\Phic{t}-\me{(\i{t}-\ip{t}-\PP\Phiu{t})}}^2\\
& \text{subject to} & & \Phic{t}+\Phiu{t}+\Phip{t}\geq 0,
\end{aligned}
\label{eq:tomography}
\end{equation}
where $\PP$ is the matrix that projects from volume into image space, i.e.,
$\PP\in \mathbb{R}^{n_p\times n_v}$; $n_p$ and $n_v$ denote the total number of
pixels and voxels, respectively. 
We use a linear image formation model, where we integrate smoke densities along a ray.
The tomography problem is solved with PD where pseudocode for is given in \myrefapp{app:reconsAlg} in \myrefalg{alg:tomo}.
When solving \myrefeq{eq:tomography} with
least-squares, it is necessary to calculate the matrix product $\PPP$. 
Because of the sparsity of the views, $n_p \ll n_v$ and 
$\PPP\in \mathbb{R}^{n_v\times n_v}$ is significantly larger and usually denser
than $\PP$. Hence, storing $\PPP$ explicitly is infeasible for larger resolutions
due to excessive memory consumption.
Therefore, we employ a CGLS solver
\cite{Ihrke:tomoFlames}, which solves the normal equations like a regular CG
solver but avoids computing $\PPP$ explicitly.

\me{For under-determined inverse problems such as our sparse tomography, 
regularizers are crucial in order to obtain smooth and realistic density
estimations as well as fluid motions.}
\nils{However, incorporating
regularizers via a square and symmetric matrix, as commonly done for regular CG solvers,
is not straightforward in CGLS. We show that regularizers
and proximal operator extensions} can be 
included in CGLS without the need for an explicit calculation of the system matrix.

A square regularization matrix $\R\in \mathbb{R}^{n_v\times n_v}$ for smoothness
and kinetic energy is very sparse in practice: \me{seven entries per row in 3D (diagonal and
neighbours for smoothness, see \myrefeq{eq:regularizers}) suffice}, and hence 
the matrix can be stored and multiplied efficiently.
Within CGLS, it is necessary to only apply $\R$ to the residual and step size calculations. 
In order to use the CGLS solver as proximal operator within a
PD optimization loop, we also need to add a weighted diagonal matrix $\sigma \mathbf{I}$
to the system matrix and the PD variable updates $\sigma \text{b}_{PD}$ to
the right-hand side.
\begin{algorithm}[t]
	\caption{Regularized CGLS}
	\label{alg:regCGLS}
	\begin{algorithmic}[1]
		\Function{\me{solveCGLSReg}}{$\PP, \R, \text{b}, \text{b}_{PD}, \sigma$}
		\State $\mathbf{r}_0 = \sigma \text{b}_{PD} - \PP^T\text{b} -\PP^T\PP x_0 - (\R+\sigma \mathbf{I}) x_0$;
		\State $\mathbf{p}_0 = \mathbf{r}_0$;
		\While{accuracy not reached}
		\State $\alpha_{k} = \norm{\mathbf{r}_{k-1}}^2 / (\mathbf{p}_{k-1}^T\PPP \mathbf{p}_{k-1} + \mathbf{p}_{k-1}^T (\R+\sigma \mathbf{I}) \mathbf{p}_{k-1})$; 
		\State $\x_{k} = \x_{k-1} + \alpha_{k} \mathbf{p}_{k-1}$;                                 
		\State $\mathbf{r}_{k} = \mathbf{r}_{k-1} - \alpha_{k} * \PPP \mathbf{p}_{k-1} + \alpha_{k} (\R+\sigma \mathbf{I}) \mathbf{p}_{k-1}; $ 
		\State $\beta_{k} = \norm{\mathbf{r}_{k}}^2 / \norm{\mathbf{r}_{k-1}}^2;  $           
		\State $\mathbf{p}_{k} = \mathbf{r}_{k} + \beta_{k} * \mathbf{p}_{k-1};     $              
		\EndWhile
		\EndFunction
	\end{algorithmic}
\end{algorithm}
This regularized CGLS solve is summarized in \myrefalg{alg:regCGLS} and used for
all tomographic density reconstructions in our framework. 

\subsection{Inflow Estimation Solver}
\label{sec:inflow}

\begin{figure}[t!]
	\begin{overpic}[width=\linewidth]{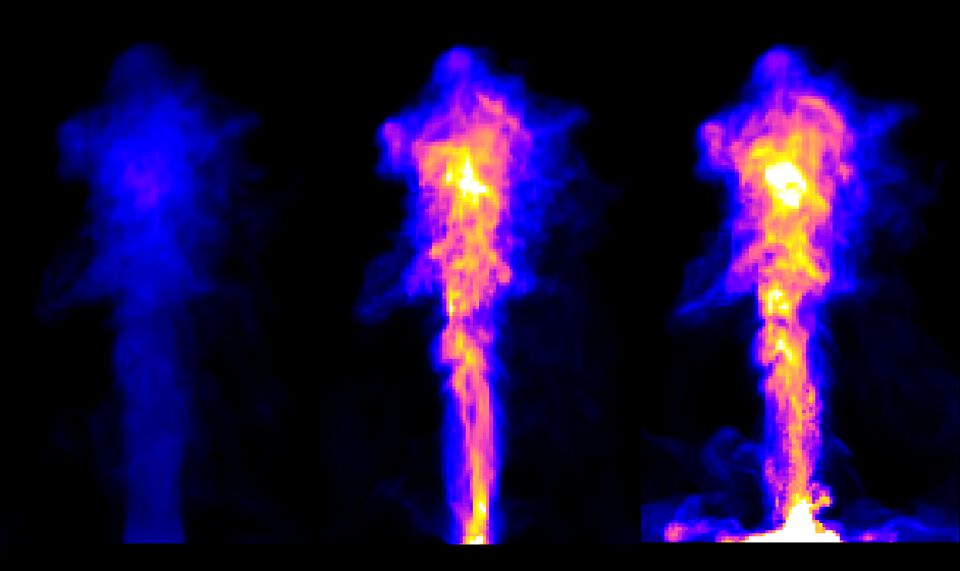}%
		\put(1,1){\textcolor{white}{\small a)}}%
		\put(34.5,1){\textcolor{white}{\small b)}}%
		\put(68,1){\textcolor{white}{\small c)}}%
	\end{overpic}
	\caption{Example reconstructions with too less density inflow a), our inflow estimation b), and too much density inflow c).}
	\label{fig:inflowIssues}
\end{figure}

Suitable boundary conditions are crucial for all physical models and are likewise
crucial for our reconstruction method.
The unseen inflow region and its boundary with the visible domain have a huge influence on the overall reconstruction quality.
Underestimating the density inflow will yield plumes that cannot fill 
the desired volume in later stages of the reconstruction, see \myreffig{fig:inflowIssues}a), while 
over-estimations can lead to strong instabilities over the course of the inverse solve, see \myreffig{fig:inflowIssues}c) for an example.
We propose an approach for solving for the correct amount of density influx by computing the unseen inflow density $\PhiI{t}$ 
considering the previous density $\Phy{t-1}$, the final velocity $\u{t}$, and the target input images $\i{t}$. Our result is displayed in \myreffig{fig:inflowIssues}b) where our reconstruction features the right amount of density to fill the plume in a stable manner.

Advecting previous and inflow density with the final velocity should result 
in a density field, the sum of which matches the amount of target smoke 
density $\Phy{t}_{\text{tar}}$ that is prescribed by the 2D input images.
The inflow density is hence used to fill the gap between current density and the target amount of density. 
%Current and target densities 
Both of them only consider the visible domain $\Gamma\setminus\Gamma_I$, i.e., not the inflow region:
\begin{equation}
\begin{aligned}
\sum_{\Gamma \setminus \Gamma_I} \mathcal{A}(\Phy{t-1}+\PhiI{t}, \u{t}) \overset{!}{=} \sum_{\Gamma \setminus \Gamma_I}  \Phy{t}_{\text{tar}}  ,
\end{aligned}
\label{eq:inflow}
\end{equation}
where $\mathcal{A}(\cdot, \u{t})$ denotes advection with the known velocity $\u{t}$ and $\Phy{t}_{\text{tar}}$ is the density target.
Based on $\u{t}$, we set up a linear system of equations 
in order to estimate the inflow densities that lead to the desired density influx;
i.e., we solve $\textbf{A}_I \PhiI{t} =$ b for $\PhiI{t}$.
In $\textbf{A}_I$, we discretize the advection operator $\mathcal{A}$ via semi-Lagrangian advection. 
The right-hand-side b contains the missing density, which should be added to the domain through the inflow solver.
Note that, as target domain, we only need to consider cells in the visible domain, which back-trace into the inflow area $\Gamma_I$, 
a region that we denote as $\Gamma_R$, as visualized in \myreffig{fig:setup2}c).
This is the only region influenced by advecting the inflow density $\PhiI{t}$.
The subset $\Gamma_R$ of $\Gamma$ is back-traced into a part of the inflow region, which is denoted as $\Gamma_{RI}$ in the following.
Hence, we set up $\textbf{A}_I$ to contain one equation for each density in $\Gamma_{RI}$ and b to contain a target value for each voxel in $\Gamma_{R}$.
We additionally ensure that each cell of the total inflow density is non-negative. 
Negative density values never occur in reality, and we found such enforcement to be crucial for 
plausible and numerically stable reconstructions.
To enforce the non-negativity constraint in conjunction with computing $\textbf{A}_I \PhiI{t}=$b, 
we solve for $\PhiI{t}$ with least-squares using smoothness and 
kinetic energy regularizers as for our main reconstruction above and employ the same convex optimization
scheme \cite{ChambollePD}.

\begin{algorithm}[bt!]
	\me{\caption{Estimate inflow, such that \myrefeq{eq:inflow} holds}
		\label{alg:estimInflow}
		\begin{algorithmic}[1]
			\Function{estimInfl}{$\i{t}, \Phy{t-1}, \u{t}$}
			\State $\Phy{t}_{\text{tar}}=\Call{reconDen}{\i{t},0}$; \Comment{target density}
			\State $\Delta \Phy{t}_{\text{tar}}=\Phy{t}_{\text{tar}} - \mathcal{A}(\Phy{t-1},\u{t})$; \Comment{residual target density}
			\For{$k\gets 1, 10$} \Comment{Prima-Dual iterations}
			\State \(\triangleright\) matrix $\mathbf{A}_I$ represents advecting cells from $\Gamma_{RI}$ to $\Gamma_R$
			\State \(\triangleright\) $\textbf{A}=\mathbf{A}_I^T \mathbf{A}_I, \mathbf{A}_I = \mathcal{A}(\PhiI{t},\u{t})|_{\Gamma_{RI}\text{ to }\Gamma_R},$
			\State \(\triangleright\) b: target missing density; $\mathcal{A}(\Phy{t-1}+\PhiI{t},\u{t})|_{\Gamma_R}\geq 0$
			\State \(\triangleright\) $\text{b}=-\mathbf{A}_I^T\text{max }\left(\Delta \Phy{t}_{\text{tar}}|_{\Gamma_R} + d,-\mathcal{A}(\Phy{t-1},\u{t})|_{\Gamma_R}\right)$
			\State $\qquad + \x^{k-1} +\sigma\y^{k-1} $
			\State $\x^k = \x^{k-1}+\sigma\y^{k-1} -\sigma \Call{solveLSE}{\textbf{A}, \text{b}}$
			\State $\z^k=\Pi_{\text{NonNeg}}( \Phy{t-1} + \z^{k-1}-\tau\x^{k})$;
			\State $\y^k=\z^k+\theta(\z^k-\z^{k-1})$;
			\EndFor
			\State $\Call{extrapolateSrcValues}{\z^k}$;
			\State return $\z^{k}$;
			\EndFunction
		\end{algorithmic}}
	\end{algorithm}
Pseudocode for the inflow estimation step is given in \myrefalg{alg:estimInflow}.
Additional details can be found in \myrefapp{app:inflow}.
We first project the target images $\i{t}$ into the volume to obtain a 3D density target $\Phy{t}_{\text{tar}}$.
The residual target density field $\Delta \Phy{t}_{\text{tar}}$, i.e., the missing density, is then given by the difference of target density and the advected previous density. 
In order to ensure the correct density influx into the visible domain $\Gamma \setminus \Gamma_I$,
the right-hand side vector $\text{b}$ contains the total missing density, i.e., $\Delta \Phy{t}_{\text{tar}}$.
As $\text{b}$ covers only a subset of the domain, i.e., $\Gamma_R$, we set each entry to the 
corresponding value in $\Delta \Phy{t}_{\text{tar}}|_{\Gamma_R}$ plus an update $d$ that 
accounts for further density discrepancies in the rest of the 
domain, i.e., $\Delta \Phy{t}_{\text{tar}}|_{\Gamma \setminus (\Gamma_R \cup \Gamma_I)}$. 
As we are solving for an inflow update to the current inflow values in $\Phy{t-1}$,  
we must ensure that $\mathcal{A}(\Phy{t-1} + \PhiI{t},\u{t})\geq0$, which 
\nils{means b$ \geq -\mathcal{A}(\Phy{t-1},\u{t})$, see line 8 of \myrefalg{alg:estimInflow}.}
We slightly extrapolate the inflow values in $\Gamma_I$ into cells that were excluded from the solve
to ensure subsequent advection steps have full access to the densities (line 13 of \myrefalg{alg:estimInflow}). 

As this solve only targets the inflow volume $\Gamma_I$, its cost is negligible compared to 
the rest of the optimization procedure. 
However, due to the non-linear nature of the overall optimization, it is a 
crucial component for obtaining realistic reconstructions.

%%% Local Variables:
%%% TeX-master: "paper"
%%% End:

%% file: 6_capture.tex
\section{Hardware Setup}\label{sec:hardware}

In addition to the algorithmic pipeline described in the previous section,
another important component of our framework is a commodity hardware setup to
capture real-world fluid flows. Our strong physics-based constraints allow us to accurately capture
complex flows with very simple hardware. In particular, complex camera
calibration and synchronization are not required. We use an insulated box, which is
heated to a chosen temperature in combination with a regular fog machine to fill
the box with a fluid that can be tracked visually. A sketch of our setup is
shown in \myreffig{fig:setup2}b. Not surprisingly, conservation of volume
holds for real fluids, and as such, it is crucial to control in- as well as
outflow to and of the box. We use two servo-controlled valves at the top and bottom of
the box, which are closed to fill the box with fog from the fog machine and
opened when initiating a capture. In practice, our heating element can yield air
temperatures of up to 60 degree Celsius ($^\circ\text{C}$) without posing safety
risks.

For camera calibration, we use a movable plate with {\em ChAr\-Uco} marker
patterns that yield a dense ray calibration for the volume right above the box.
After calibration, we cover the box and the background with a diffuse black cloth in order to
maximize contrast of the visible fog. To record the fog, we use a set of five
Raspberry Pi computers with attached cameras mounted on microphone stands. In
total, the whole capture setup consists of hardware that is available for less than
1100 USD. Thus, in contrast to previously proposed hardware setups in graphics
\cite{Hawkins:timeVary,Xiong:2017}, our setup\footnote{Technical details will be
  published together with our data set.} can be recreated with a very moderate
investment. Furthermore, other algorithms \cite{Gregson:2014} typically require
larger numbers of synchronized and carefully calibrated cameras.

%%% Local Variables:
%%% TeX-master: "paper"
%%% End:

%% file: 4_results.tex
\section{Evaluation}\label{sec:eval}

\begin{figure*}[t!]
	\begin{subfigure}[t]{0.315\textwidth}
		\includegraphics[trim={5 0 1355 0},clip,width=0.47\linewidth]{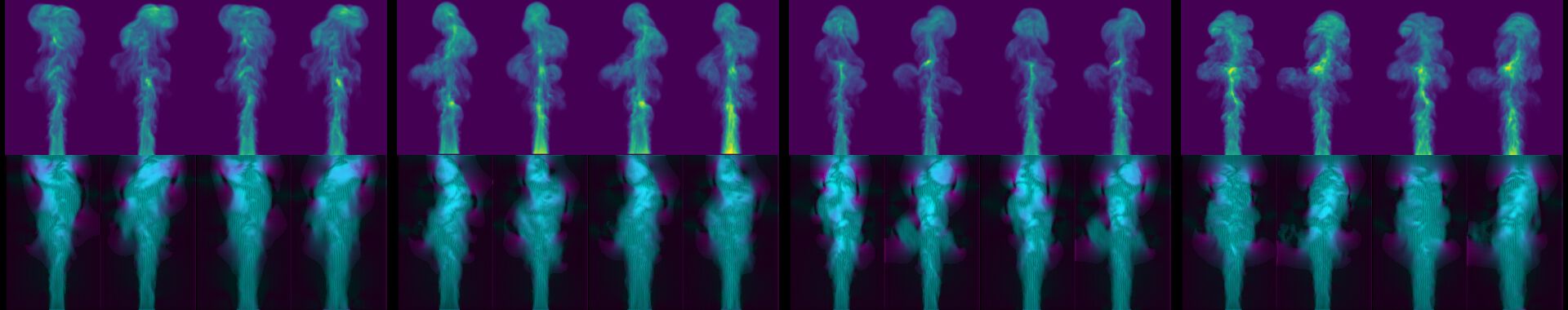}
		\includegraphics[trim={185 0 1175 0},clip,width=0.47\linewidth]{images/synthEval.jpg}	%\vspace{0.3cm}	
		\caption{\scriptsize{Base simulation, front, \mbox{$t=170$}.}}
		\label{fig:evalA}
	\end{subfigure}	
\hfill
	\begin{subfigure}[t]{0.315\textwidth}
		\includegraphics[trim={365 0 995 0},clip,width=0.47\linewidth]{images/synthEval.jpg}
		\includegraphics[trim={545 0 815 0},clip,width=0.47\linewidth]{images/synthEval.jpg} %\vspace{0.3cm}
		\caption{\scriptsize{Larger inflow area, front, \mbox{$t=145$}.}}
		\label{fig:evalB}
	\end{subfigure}	
\hfill
	\begin{subfigure}[t]{0.315\textwidth}
		\includegraphics[trim={815 0 545 0},clip,width=0.47\linewidth]{images/synthEval.jpg}
		\includegraphics[trim={995 0 365 0},clip,width=0.47\linewidth]{images/synthEval.jpg} %\vspace{0.3cm}
		\caption{\scriptsize{$50~\%$ more buoyancy, front, \mbox{$t=122$}.}}
		\label{fig:evalC}
	\end{subfigure}	
\hfill
	\begin{subfigure}[t]{0.315\textwidth}
		\includegraphics[trim={1175 0 185 0},clip,width=0.47\linewidth]{images/synthEval.jpg}
		\includegraphics[trim={1355 0 5 0},clip,width=0.47\linewidth]{images/synthEval.jpg} %\vspace{0.3cm}
		\caption{\scriptsize{High-re\-so\-lu\-tion, front, \mbox{$t=155$}.}}
		\label{fig:evalD}
	\end{subfigure}	
\hspace{0.3cm}
	\begin{subfigure}[t]{0.315\textwidth}
		\includegraphics[trim={95 0 1265 0},clip,width=0.47\linewidth]{images/synthEval.jpg}
		\includegraphics[trim={275 0 1085 0},clip,width=0.47\linewidth]{images/synthEval.jpg}
		\caption{\scriptsize{Base simulation, side, \mbox{$t=170$}.}}
		\label{fig:evalAS}
	\end{subfigure}	
	\hfill
	\begin{subfigure}[t]{0.315\textwidth}
		\includegraphics[trim={455 0 905 0},clip,width=0.47\linewidth]{images/synthEval.jpg}
		\includegraphics[trim={635 0 725 0},clip,width=0.47\linewidth]{images/synthEval.jpg}
		\caption{\scriptsize{Larger inflow area, side, \mbox{$t=145$}.}}
		\label{fig:evalBS}
	\end{subfigure}	
	\hfill
	\begin{subfigure}[t]{0.315\textwidth}
		\includegraphics[trim={725 0 635 0},clip,width=0.47\linewidth]{images/synthEval.jpg}
		\includegraphics[trim={905 0 455 0},clip,width=0.47\linewidth]{images/synthEval.jpg}
		\caption{\scriptsize{$50~\%$ more buoyancy, side, \mbox{$t=122$}.}}
		\label{fig:evalCS}
	\end{subfigure}	
	\hfill
	\begin{subfigure}[t]{0.315\textwidth}
		\includegraphics[trim={1085 0 275 0},clip,width=0.47\linewidth]{images/synthEval.jpg}
		\includegraphics[trim={1265 0 95 0},clip,width=0.47\linewidth]{images/synthEval.jpg}
		\caption{\scriptsize{High-re\-so\-lu\-tion, side, \mbox{$t=155$}.}}
		\label{fig:evalDS}
	\end{subfigure}	
	\caption{Four different synthetic smoke simulations. Each with ground truth density (left) and reconstructed density (right), 
	with ground truth velocity (lower left) and reconstructed velocity (lower right), for front and side views.
		The high-re\-so\-lu\-tion simulation and reconstruction in d) and h) are downsampled to facilitate visual comparison. 
		We use our real-world calibration data to generate the synthetic input images and use our inflow estimation for reconstruction. 
		Across varying simulation parameters as well as resolution, our reconstructions recover both density and motion accurately.}
	\label{fig:eval}
\end{figure*}

To evaluate the accuracy of our pipeline, we use a series of simulated data sets
for which we have known ground truth values available for all quantities. 
We investigate the reconstruction accuracy for varied parameters and at varying resolutions. 
The perturbed parameters mimic the unknown behavior 
of real-world setups and, thus, can be used to assess the robustness of our reconstruction algorithm 
in practical settings.
For each scenario, we vary the noise in the inflow density of the simulation five times such that we obtain five different instances for each synthetic case.
We reconstruct all five instances and calculate the mean and standard deviation of peak signal-to-noise ratio (PSNR) for differences in density, velocity, and images. 

\subsection{Reconstruction Accuracy}

\begin{figure*}[]
	\captionsetup[subfigure]{aboveskip=-5pt,belowskip=0pt}
	\subcaptionbox{Base simulation}      {\includegraphics[width=0.4\linewidth]{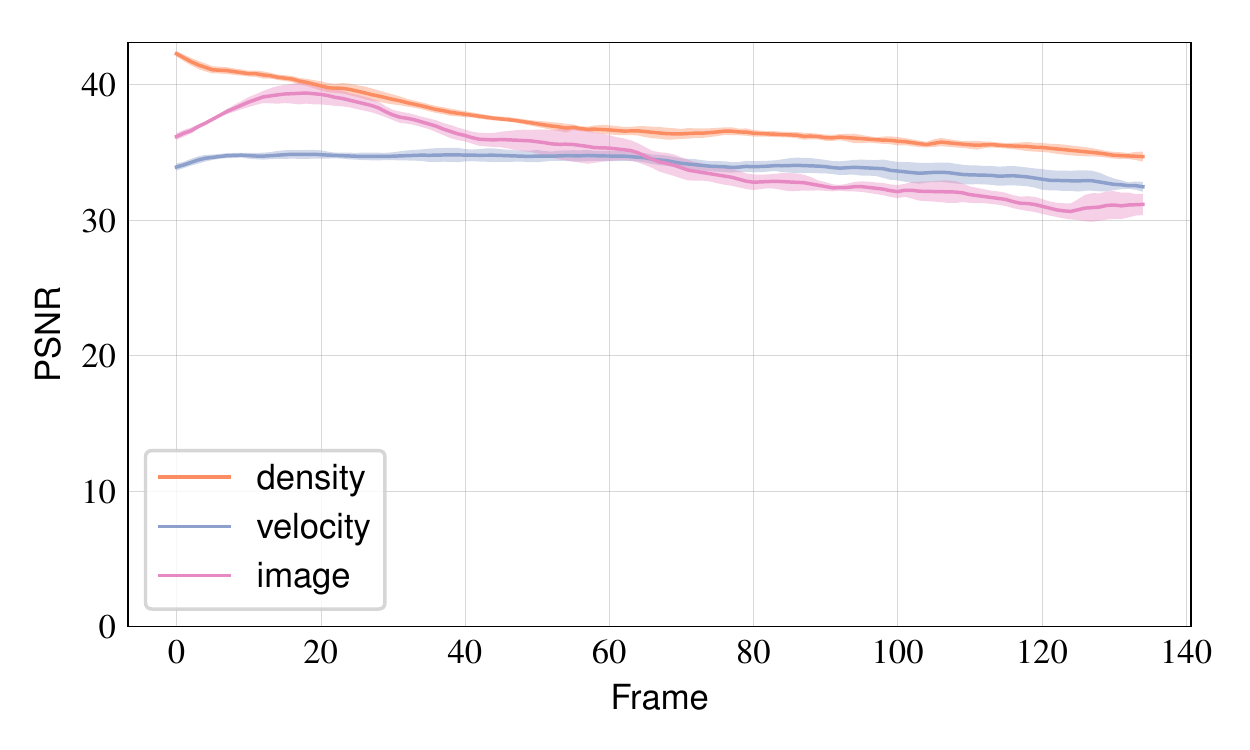}}
	\subcaptionbox{Increased inflow area}{\includegraphics[width=0.4\linewidth]{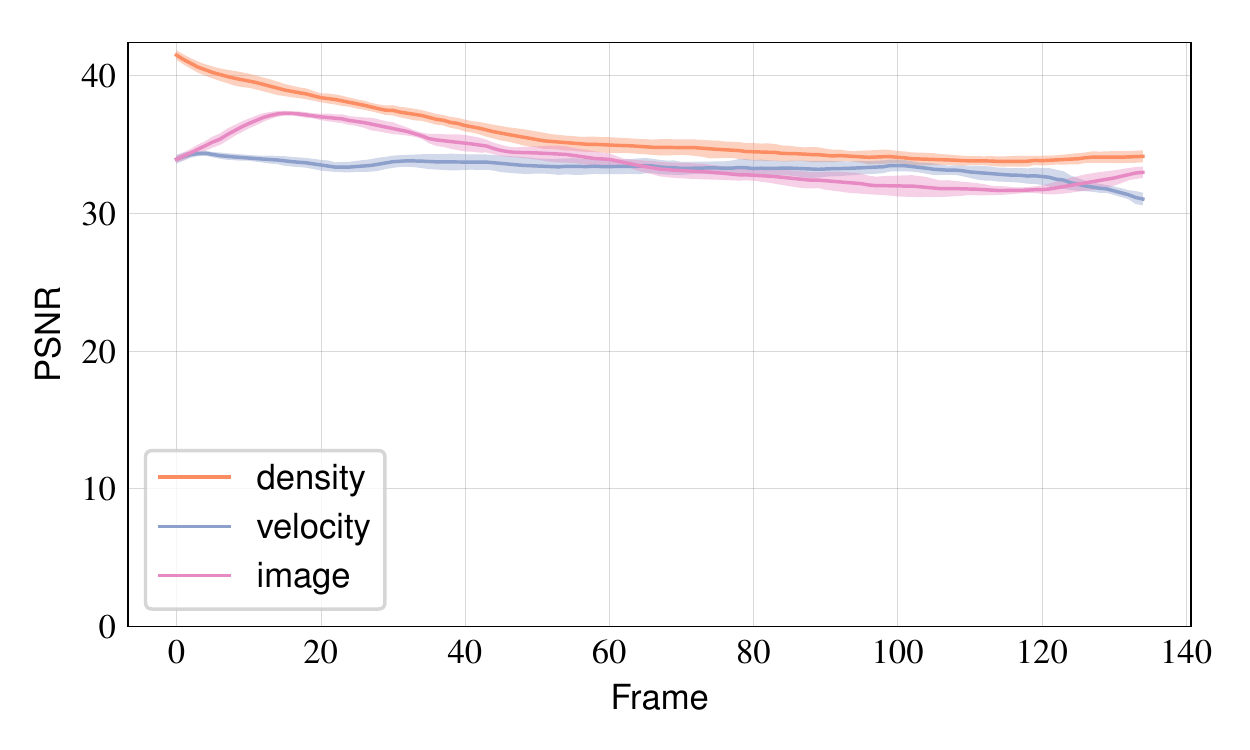}}\\
	\subcaptionbox{Increased buoyancy}   {\includegraphics[width=0.4\linewidth]{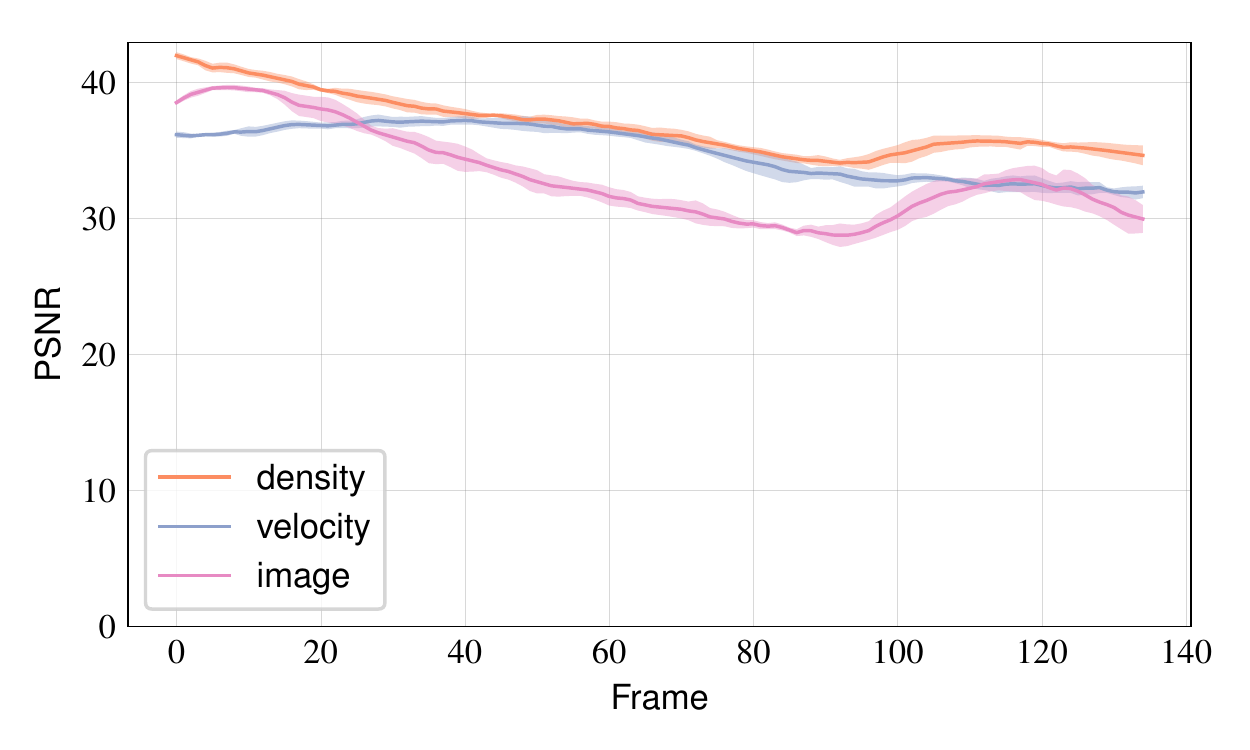}}
	\subcaptionbox{Increased resolution} {\includegraphics[width=0.4\linewidth]{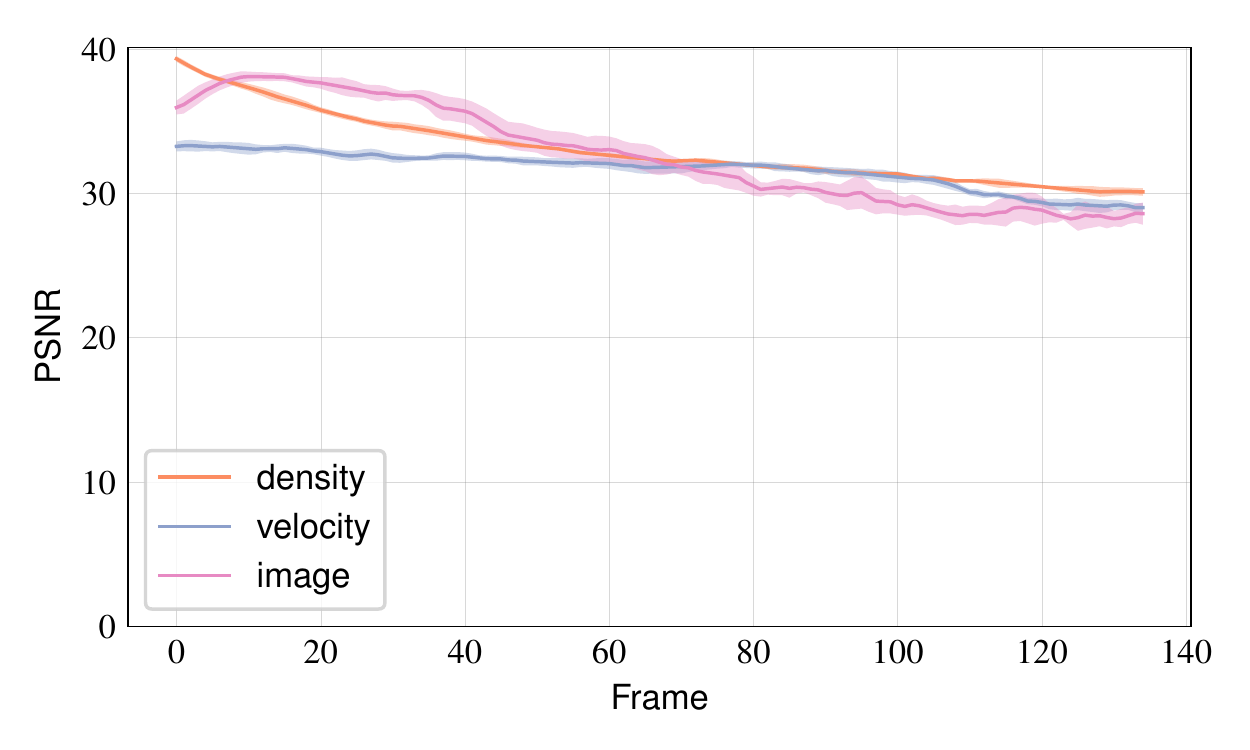}}
	%\subcaptionbox{Density}{\includegraphics[width=0.32\linewidth]{images/psnry0-synt0}}
	%\subcaptionbox{Velocity}{\includegraphics[width=0.32\linewidth]{images/psnry0-synt1}}
	%\subcaptionbox{Images}{\includegraphics[width=0.32\linewidth]{images/psnry0-synt2}}
	\caption{PSNR values of differences in density, velocity, and images for our four synthetic reconstructions in \myreffig{fig:eval} with five different simulation and reconstruction instances each. We show the mean and standard deviation of PSNR values across the five instances.}
	\label{fig:psnrSynthSep}
\end{figure*}

First, we simulated a plume of hot smoke with the Boussinesq approximation at a resolution of
$100 \times 177 \times 100$ and reconstructed the data set with the 
same resolution using our full algorithm.
% discuss separately later
The plumes were rendered with our raycaster from five
virtual views with real-world camera calibrations in line with our hardware
setup. We then reconstructed the flow based on these images without using any
additional information such as the ground truth inflow. The inflow was likewise
computed with our estimation algorithm. 
This baseline comparison achieves a very high accuracy with averaged PSNR values of $37.3, 34.1$, and $34.6$ for density, velocity and image differences, respectively, across all five instances, see \myreffig{fig:psnrSynthSep}a).
A visual example can be found in \myreffig{fig:eval}a,e). Note that even the side view, 
which is heavily under-constrained in terms of visual observations, is reconstructed very accurately.

To evaluate the robustness of our method, we
vary the size of the inflow source or 
increase the buoyancy force by $50\%$. Thus, in combination with the test above, these three 
tests yield distinct data points in the space of possible buoyant plume parameters. 
The resulting simulations exhibit significantly different flow behavior as observed 
in \myreffig{fig:accGraph}b,c). Despite these variations, our algorithm very accurately reconstructs 
the different flows in terms of both visual density and flow motion (bottom row of \myreffig{fig:eval}).
The error measurements provide PSNR values comparable to the earlier test, i.e., an average of $35.7, 33.3, 33.8$ and $36.8, 34.9, 33.1$ for density, velocity, and image differences, respectively, across all five instances, see \myreffig{fig:psnrSynthSep}b,c).
These tests indicate that our method is capable of accurately reconstructing a variety of 
different buoyant flows.
For our scenario with increased buoyancy, a slight drop in PSNR values is observed around $t=95$.
Due to the increased buoyancy, the smoke plumes rise much faster and, as such, leave the domain around $t=95$.
In order to arrive at a setup that
resembles the challenging real-world conditions, we also used a reduced resolution of $100 \times 177 \times 100$ for the reconstruction of a 
synthetic high-re\-so\-lu\-tion simulation with a doubled resolution of $200 \times 354 \times 200$.
Despite the inherently different resolution used for reconstruction, 
the result of our algorithm closely
matches the ground truth for both density and velocity (\myreffig{fig:eval}d,h). 
The averaged PSNR values are $33.0, 31.6,$ and $32.6$ for density, velocity, and image differences, respectively, across all five instances, see \myreffig{fig:psnrSynthSep}d).

While our reconstructions exhibit slightly less overall density, 
it accurately captures the intricate shapes of the complex reference flows as shown visually in \myreffig{fig:eval} and measured through PSNR values, see \myreffig{fig:accGraph}. 
All four tests with five instances each robustly produce similar PSNR values for density, velocity, and images over multiple time steps. 
As reconstruction proceeds, the PSNR values slightly drop, but they yield high overall averages across all five instances. 
The decrease of the PSNR values over time can be explained by the flow becoming more complex over time and by occupying a larger volume of the domain.
From these tests, we conclude that, with our strong physics-based optimization, five
camera views are sufficient to constrain a simulation to recreate realistic flow motion according to given input images.

\subsection{Alternative Methods}

\begin{figure}
	\begin{overpic}[width=1.\linewidth]{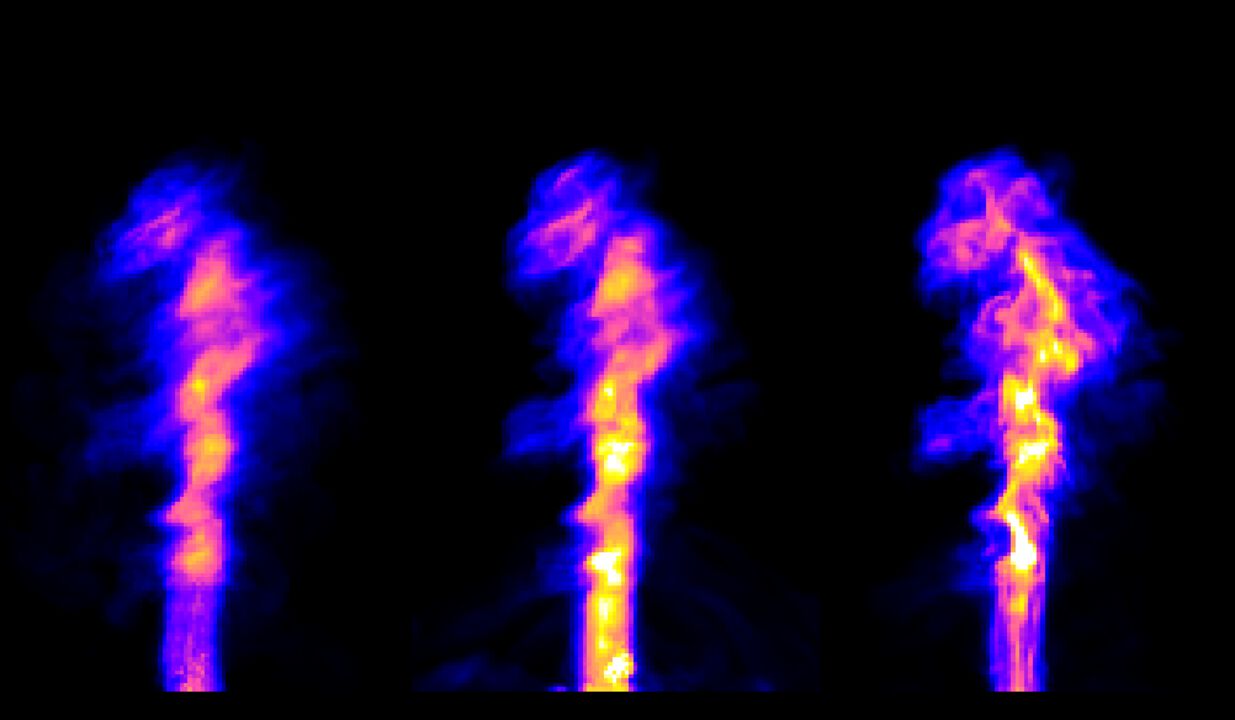}%
		\put(2,54){\textcolor{white}{\footnotesize a) Tomo + OF: $V_a$}}%
		\put(35,54){\textcolor{white}{\footnotesize b) Simplified: $V_b$}}%
		\put(68,54){\textcolor{white}{\footnotesize c) Ours}}%
	\end{overpic}
	\caption{\nils{A visual comparison of three methods
			in terms of reconstructed densities: a) a divergence-free optical flow reconstruction based on tomographic densities from previous work, b) a simplified version of our solver, and c) our full method.
			Our full algorithm is needed to obtain a realistic and detailed flow field without artifacts.}
	}
	\label{fig:relWork}
\end{figure}

\begin{figure}
	\centering
	\includegraphics[width=\linewidth]{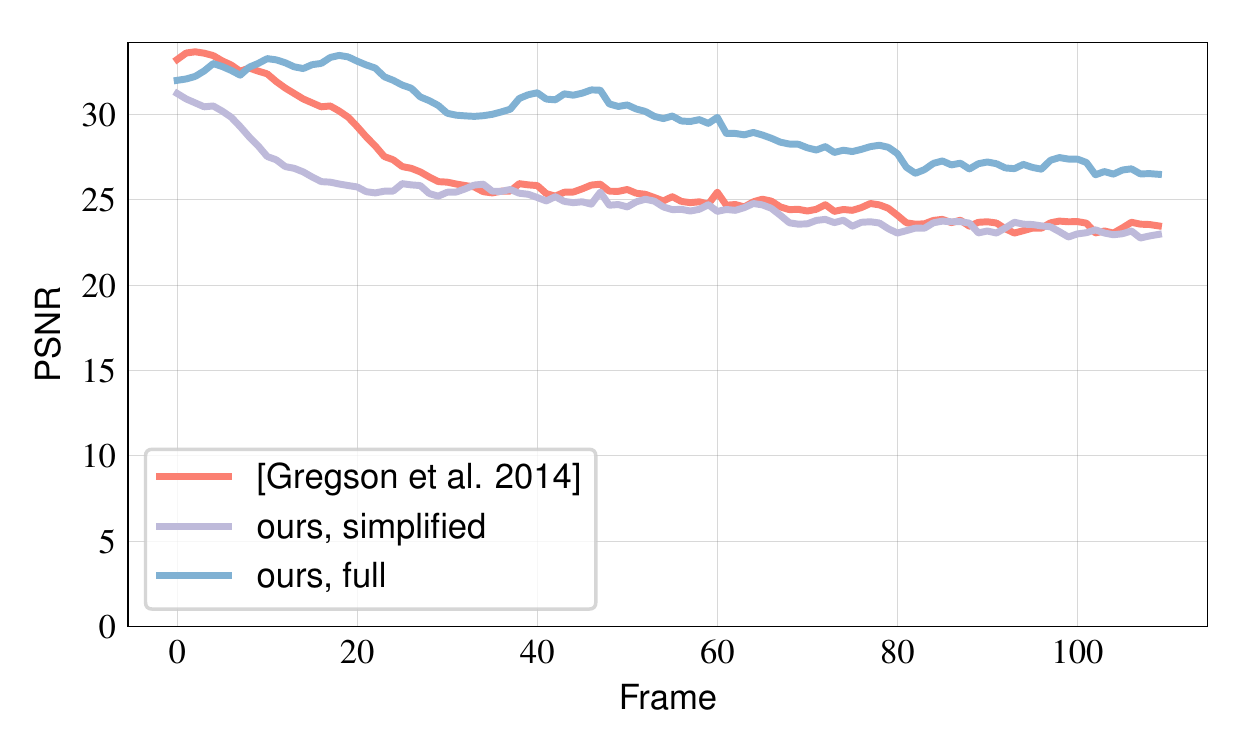}
	\caption{PSNR of image differences for $V_a$ (tomography plus 3D optical flow), $V_b$ (a simplified version of our approach), and our method.}
	\label{fig:psnrAltMethods}
\end{figure}

In addition, we evaluate the gains in accuracy of our approach 
with respect to a flow reconstruction method from previous work
and a simplified variant of our algorithm.
The former approach, denoted as $V_a$ in the following, 
first computes volumetric densities through tomography and then applies 3D optical flow that is constrained to be divergence-free (in line with Gregson et al. \shortcite{Gregson:2014}). 
The second variant, denoted as $V_b$, uses our algorithm, but omits solving iteratively for an advection-aware 
residual density as outlined in \myrefeq{eq:problem_statement}, and instead uses a single delta density obtained from a separate tomography solve.
The averaged PSNR values for image differences are $25.8, 25.0,$ and $29.2$ for $V_a, V_b$, and our method, respectively.
These measurements show that our method reproduces the target input images with the highest accuracy among these methods, which is indicated by the per-time step PSNR measurements in \myreffig{fig:psnrAltMethods}. 
Although our method outperforms the others regarding error in the image space, there is more to consider as 2D error measurements cannot fully evaluate the quality of the volumetric reconstructions. 
Therefore, a qualitative evaluation in terms of densities is visualized in \myreffig{fig:relWork}.	
As shown in \myreffig{fig:relWork}a), the version $V_a$ exhibits the typical stripe artifacts from tomographic reconstructions
leading to an overall lack of detail and clearly visible artifacts.
Variant $V_b$ fares better as shown in \myreffig{fig:relWork}b) but likewise lacks detail in the side view. Additionally, 
the reconstruction cannot keep up with the observations such that only two thirds of the overall length can be reconstructed.
Our full algorithm shown in \myreffig{fig:relWork}c) reconstructs the full sequence and develops natural, 
fluid-like behavior without tomography artifacts.
These comparisons highlight that our iteratively re-computed density change is a density field that lies 
within the image formation null space and \me{matches fluid motions} significantly better than the updates computed by previous work. 
This improved quality is caused by the \me{enhanced} physical constraints which \me{lead to} a more realistic solution 
from the aforementioned null space. In this way, our method is able to produce more natural behavior and
density configurations without the typical tomography artifacts \me{and is especially suitable for under-constrained problems.}

\nt{\subsection{Performance}}
We also employ this setup to evaluate the performance of our optimized CGLS
regularization. Here, a regular CG solver requires 535 seconds per tomography
solve on average, while our CGLS solver achieves the same residual accuracy with
only 69 seconds on average. While, for the regular CG, 182s are spent on the
matrix construction, only 13s are required for CGLS. Thus, instead of being a
bottleneck, the tomographic reconstruction becomes a smaller part of the overall
run time, which was 809s and 350s on average per time step for CG and CGLS,
respectively. Here, the performance of the CG version is indicative for the
run times of previously proposed methods \cite{eckert18}.

In addition, the full CG solver requires approximately 15 GB of
memory to store the tomography matrix $\PPP$, while the CGLS solver needs only
2.5 GB. As both memory and computational requirements grow super-linearly for
larger resolutions, the regularized CGLS is crucial for obtaining reasonable run
times for larger resolutions.

\section{Data Set} \label{sec:dataset}

\begin{figure}[]
	\includegraphics[width=\linewidth]{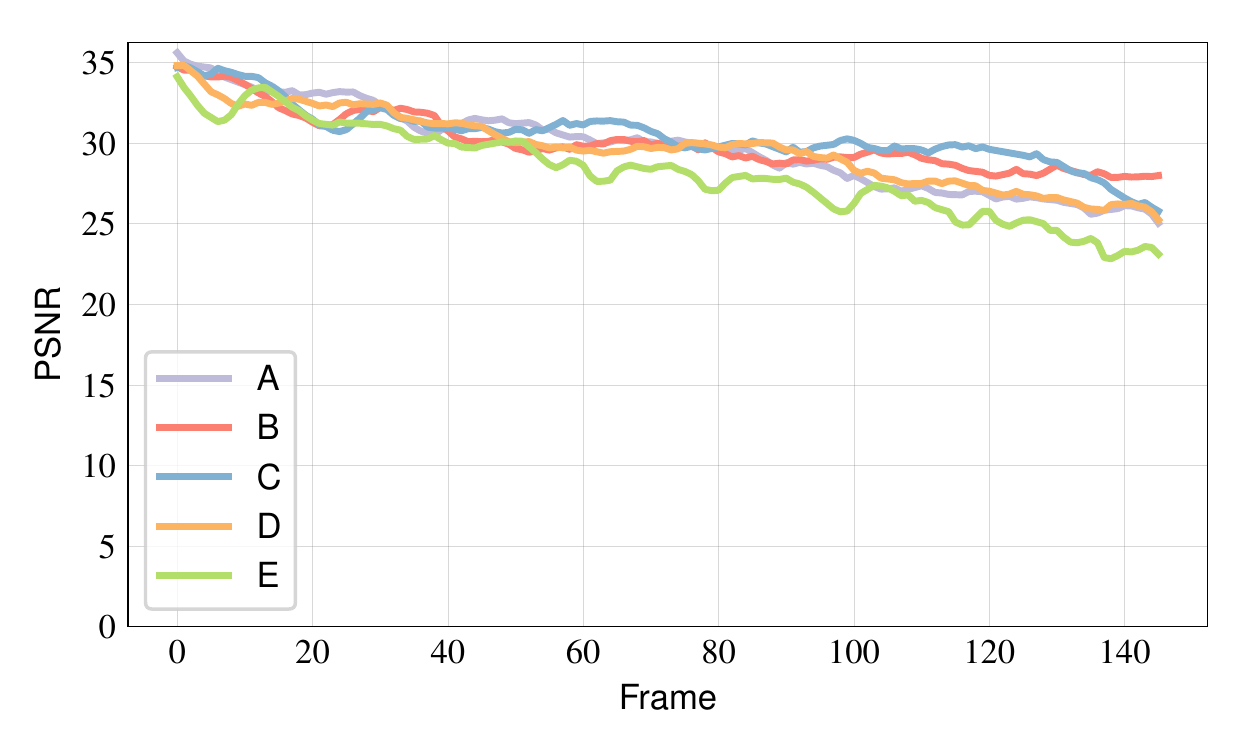}
	\caption{PSNR values of image differences for five different real-world reconstructions.}
	\label{fig:accGraph}
\end{figure}

We now give an overview of the {\em ScalarFlow} data set, which is available online.
The 
data set contains 100 flow reconstructions, each with a resolution of
100$\times$177$\times$100 and with at least 150 time steps (ca. 26 billion
voxels of data in total). 
During the capture experiments, we used a temperature of $34^\circ~\text{C}$ as higher
temperatures tend to move the visible flow too far out of the calibrated volume
before producing interesting instabilities. This produces flow velocities of ca. 0.3 to
0.4$~\frac{\text{m}}{\text{s}}$ on average, once the fog plumes have accelerated. Naturally, the initial
velocity for all captured flows is close to zero. Relative to the size
of our calibrated capture volume of 0.9~m, this yields Reynolds numbers of up to ca. $5400$.
Thus, the captured flows transition to turbulence during the later stages of each capture.
In our reconstructions, we assume a viscosity of $1.516 \cdot 10^{-5}~\frac{\text{m}^2}{\text{s}}$ for the fluid,
and we record the rising plumes with $60$~fps.

\begin{figure*}[tb]
	\includegraphics[width=.49\linewidth]{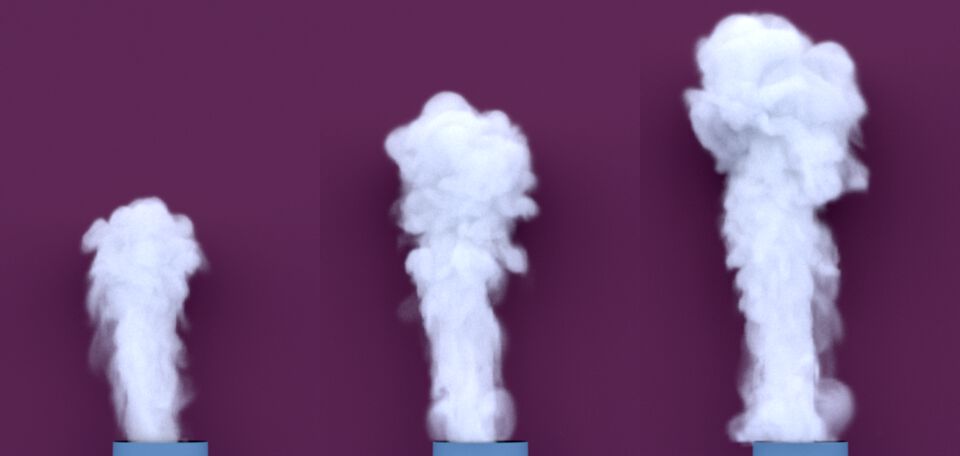}\hfill
	\includegraphics[width=.49\linewidth]{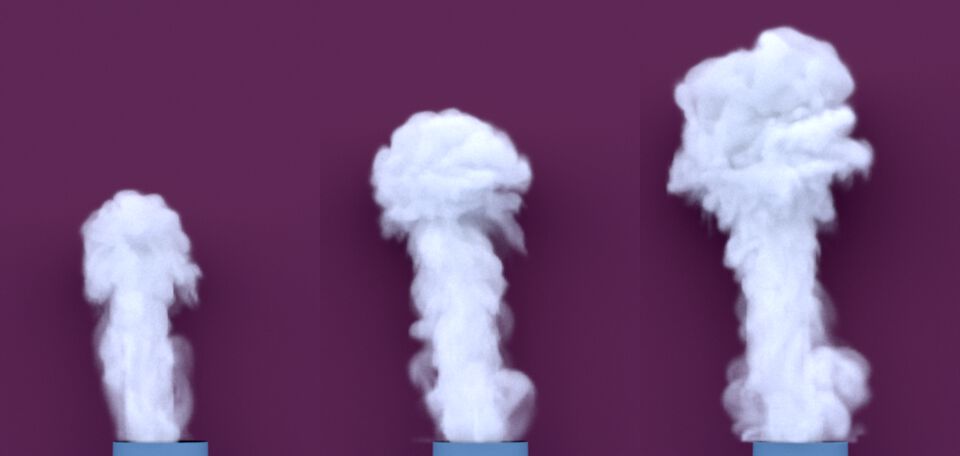}\\
	\includegraphics[width=.49\linewidth]{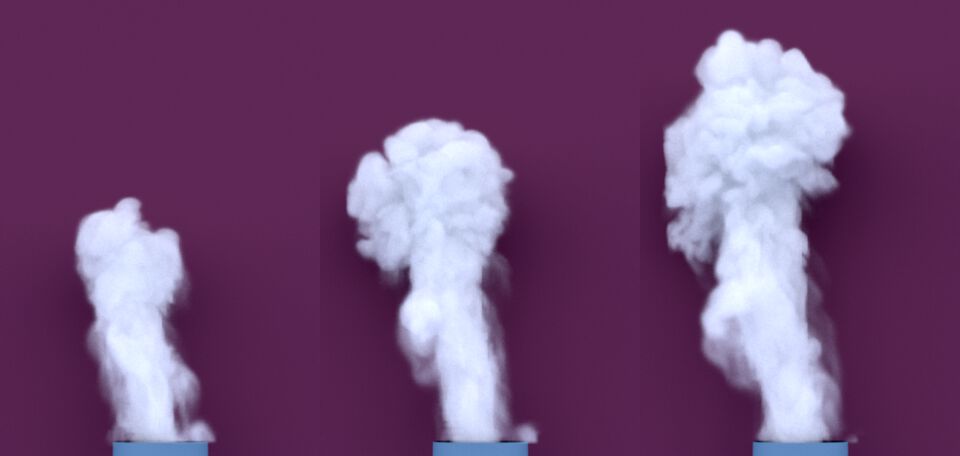}\hfill
	\includegraphics[width=.49\linewidth]{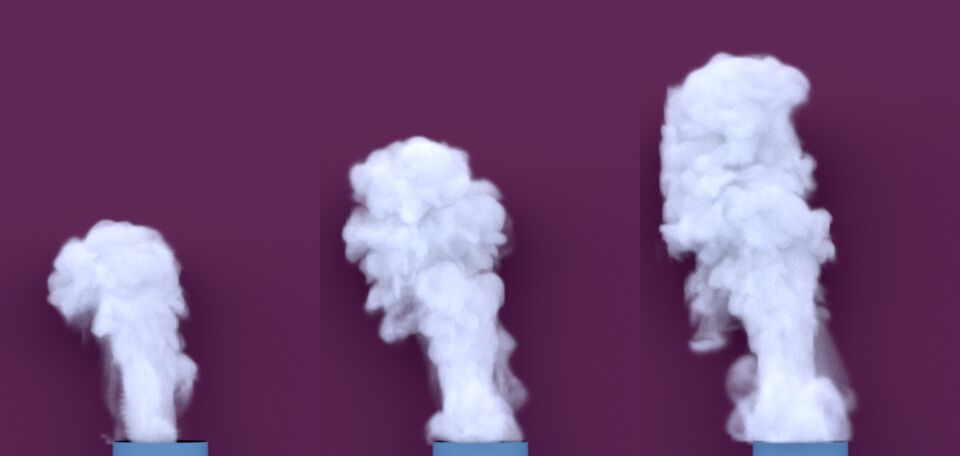}
	\caption{
		\nt{Four captured sequences shown in terms of \me{thickly rendered} reconstructed densities.
			Here, we visualize the captured plumes in a way that is very different to their 
			real-world counterparts. This rendering style highlights the amount of small scale 
			detail contained in our reconstructions.}
		\label{fig:mitsuba} }
\end{figure*}
Additional density and velocity visualizations for a selection of 20 reconstructions are
shown in \myreffig{fig:collection} in \myrefapp{app:addRecon}.
These visualizations show that our data set
contains largely similar plume motions, which is important as our goal is to
thoroughly sample a chosen space of physical behavior. However, despite the
overall similarity, the captured flows contain a variety of interesting and
natural variations, such as secondary plumes that form at various stages and
separate from the main plume. 

The volumetric velocity and density data contained in this collection can be
flexibly used, e.g., for evaluations, re-simulations, data-driven applications,
or visualizations. Example renderings of a sub-set of the density data are shown
in \myreffig{fig:mitsuba}. 
As for our comparison to alternative methods, we evaluate the accuracy of our real-world
reconstructions by computing the PSNR between the captured images and the rendered
reconstructed densities. As shown in \myreffig{fig:accGraph}, the
images match the captured images very well with average PSNRs between 
$27.4$ and $29.7$. All five reconstructions robustly result in similar error ranges.

\section{Perceptual Evaluation}\label{sec:app}

As a first exemplary application of our data set, we show a perceptual
evaluation of different simulation methods and resolutions for buoyant smoke 
clouds via user
studies. To achieve robust and reliable evaluation results in these studies, a
key requirement is to have reference data that can serve as ground truth
and to have a large number of available data variants of the same phenomenon 
in order to increase robustness. 
For our user studies, we adopt the two-alternative forced choice (2AFC) design
\cite{fechner1860} and compute scores with the Bradley-Terry model
\cite{hunter2004,um2017perceptual}. 
\ku{We recruited 189
participants from 48 countries via crowd-sourcing, where each participant
answered the randomized questions twice. In total, we had 100 answers per
question to obtain statistically relevant result.}

We first consider a selection of fluid solvers that are well established in the
community in order to evaluate how closely their results resemble a real fluid.
To this end, we selected four representative methods: semi-Lagrangian advection
\cite{Stam1999}, MacCormack \cite{Selle:2008:USM}, advection-reflection
\cite{zehnder2018}, and wavelet turbulence \cite{Kim:2008:wlt} as a
representative of up-res methods. While these methods have been compared
visually in the corresponding publications, to the best of our knowledge, no
evaluation of these methods in comparison to a real-world reference exists.

\begin{figure}[tb]
  \includegraphics[width=.99\columnwidth]{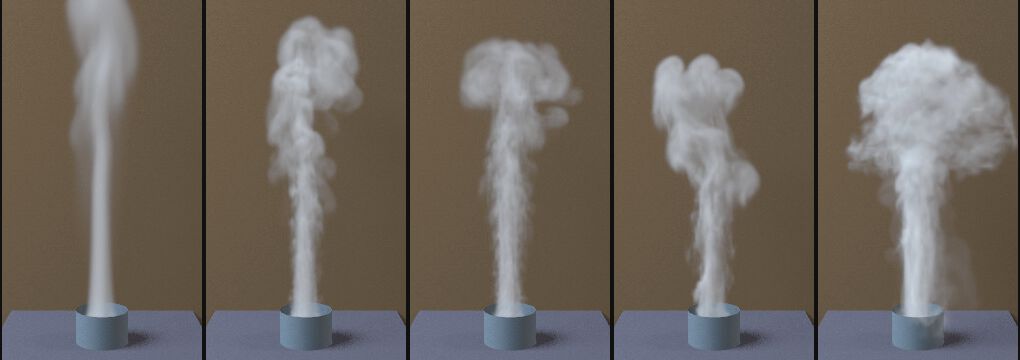}
  \caption{Example frames of user study videos for four different simulation
    methods, and one ScalarFlow data set. F.l.t.r.: Semi-Lagrangian, MacCormack,
    advection-reflection, wavelet turbulence, and ScalarFlow data.}
  \label{fig:plume-methods}
  \vspace{0.2cm}
  \includegraphics[width=.99\columnwidth]{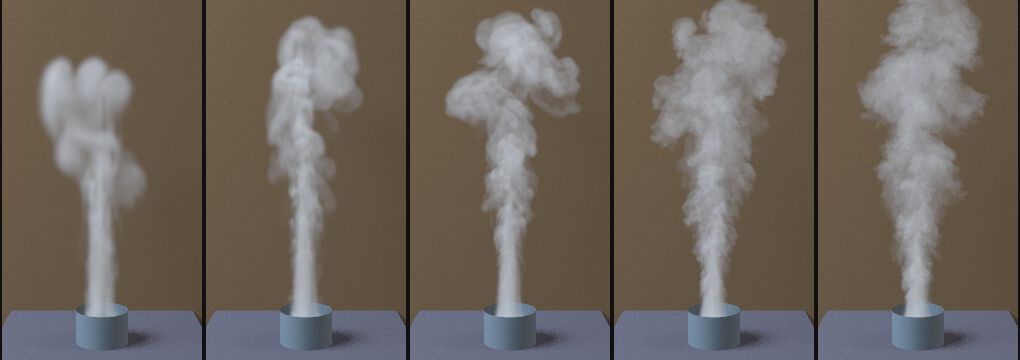}
  \caption{Example frames of user study videos for increasing resolutions. The
  	simulations with MacCormack advection for 1$\times$, 2$\times$, 4$\times$, 8$\times$, and 12$\times$
  	are shown from left to right, respectively.}
  \label{fig:plume-resolutions}
\end{figure}

We implemented all simulation methods in the same solver framework
\cite{mantaflow} such that all four simulation variants use the same initial
conditions and a resolution of 100$\times$177$\times$100. The base resolution of
the wavelet turbulence version was halved such that the synthetic turbulence can
be added with a two times up-sampling. \myreffig{fig:plume-methods} shows
example frames of the four different simulations as well as one of our
reconstruction data set.

\begin{figure*}[tb]
  \centering
  \captionsetup[subfigure]{aboveskip=1pt,belowskip=0pt}
  \subcaptionbox{Different methods\label{fig:us-method}}{\includegraphics[width=.43\linewidth]{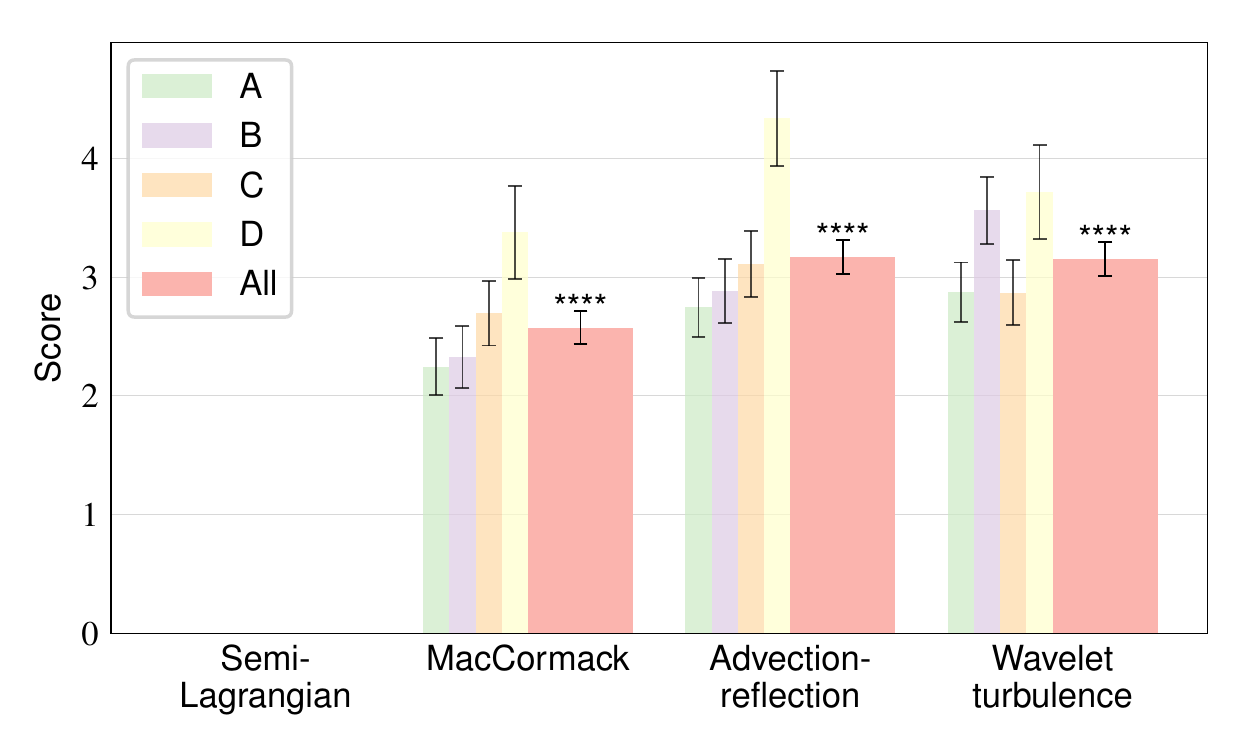}}
  \subcaptionbox{Different resolutions\label{fig:us-resolution}}{\includegraphics[width=.43\linewidth]{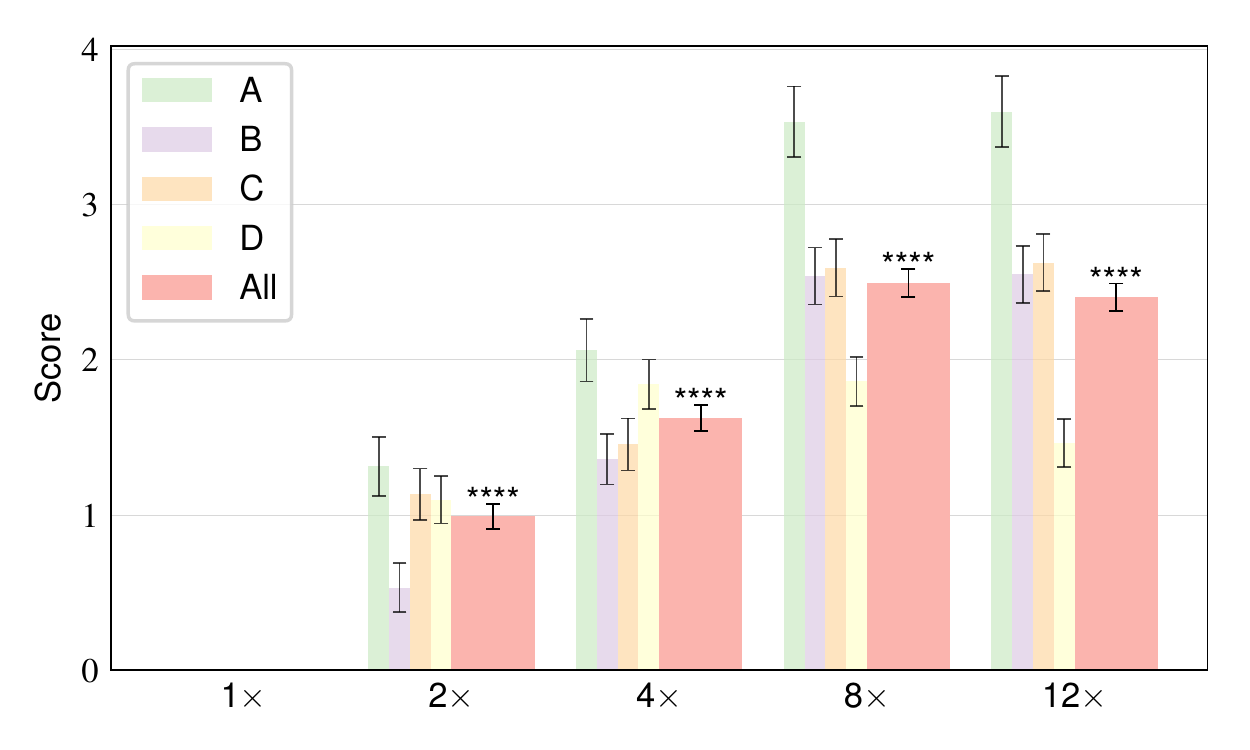}}
  \caption{\ku{Evaluations of different simulation methods and different resolutions
    via user studies. (\subref{fig:us-method}) Multiple evaluations (i.e., A, B,
    C, and D) of the same four methods are shown as well as their combined
    evaluation (i.e., All). In (\subref{fig:us-resolution}) the evaluation of
    different resolutions is shown. ****$P<0.0001$.}}
  \label{fig:userstudy-plume}
\end{figure*}

The result of our 2AFC user study with the different methods is summarized in
\myreffig{fig:userstudy-plume}\subref{fig:us-method}. Our study shows which
method resembles a real smoke cloud more closely. For instance, our evaluations
show a chance of \ku{64\%} that viewers prefer the advection-reflection version over
the MacCormack result in comparison to our reconstruction.

It is typically crucial to have full control over the setup of the user studies.
With the available 3D density of our data set, we are able to flexibly design our psychophysical studies
with custom background, smoke color, rendering style, and viewing angles.
Furthermore, thanks to the variety of reconstructions in the ScalarFlow data set, we can
evaluate the aforementioned methods in comparison to multiple real flows.
Despite their different behaviors, a range of salient flow features 
appears across these smoke clouds. In this way, we can determine the viewer's
preferences for a wider range of natural flow behavior, and we can ensure that
a singular result is not an outlier. Conducting four additional user studies
for the same method yet with different reference data, we found that the
results of these studies were all highly correlated, \ku{$\rho\simeq0.98$
$(P\simeq0.02)$ on average}, where $\rho$ and $P$ are the correlation coefficient and its
p-value, respectively. This indicates that our evaluation results are stable.
The joint preferences of our participants, computed across all four studies, are
shown in red in \myreffig{fig:userstudy-plume}\subref{fig:us-method}.
Interestingly, the relatively old, procedural wavelet turbulence method
\ku{exhibits a performance that is comparable to the advection-reflection solver}
\nils{ and slightly outperforms the MacCormack scheme}.

In addition to the different simulation methods, we further investigate the influence of
simulation resolution. As graphics solvers typically do not explicitly add
viscosity, they rely on unknown amounts of numerical viscosity in order to
achieve a realistic look. With this study, our goal was to investigate how much
numerical viscosity is actually necessary to realistically simulate a 
plume of buoyant smoke approximately one meter high. Here, we focus on the MacCormack method as a
\emph{de facto} standard advection scheme, which we use to simulate five
different resolutions. Starting with a resolution of 50$\times$88$\times$50 as
the base resolution (i.e., 1$\times$), we increase the resolution by 2$\times$,
4$\times$, 8$\times$, and 12$\times$, respectively, arriving at
600$\times$1062$\times$600 for the most finely resolved version with 12$\times$.
Note that we use a scaling factor of $1.77$ for the domain's height.
Example frames are shown in \nt{the supplemental material in} \myreffig{fig:plume-resolutions}.

\ku{Our user study, summarized in
\myreffig{fig:userstudy-plume}\subref{fig:us-resolution}, shows that the
8$\times$ and 12$\times$ resolutions are those} \nils{
that are considered to be closest to our captures.
Interestingly, there is a noticeable variance in the evaluations across
the different data sets. For version {\em D}, the 8$\times$
simulation is even considered to be closer to the capture than the 12$\times$ simulation.
This behavior illustrates the need for a large number of data 
sets in order to ensure a robust evaluation.
This result additionally shows that large resolutions are required
in numerical simulations to perceptually match the behavior of real-world
smoke clouds at a scale of ca. 90~cm. Consequently, larger real-world clouds
would require even larger resolutions.}

\section{Limitations}\label{sec:limitations}

Our reconstruction algorithm is a non-linear optimization procedure, and as
such, is not guaranteed to converge. However, this is a limitation that our method shares
with all previous work in this area. We found the algorithm to be stable in
practice, but substantial changes in the data, like speed or brightness, can
lead to diverging reconstruction runs. Here, a promising avenue for future work
would be a \nils{further improved estimation of the inflow velocity and an 
automated adjustment of the reconstruction parameters}, which could lead to even 
more accurate reconstructions.

In practice, the ambient air motion of our capture stage makes it difficult to
fine-tune the direction of the plumes. 
Right now, applications of our data set have to include slight variations in terms of the
overall plume direction. However, the data could be clustered with respect to
average motions of the plume, and future extensions of the data set could yield
enough samples such that individual directions of motion could be targeted.

In addition, we currently rely on a linear image formation model. While we found
this to yield very good results for the relatively thin clouds of our fog
machine, the linear image formation model would not be directly applicable to
denser volumes.

%%% Local Variables:
%%% TeX-master: "paper"
%%% End:

%% file: 5_conclusion.tex
\section{Conclusions}\label{sec:conclusion}

We have shown that it is possible to accurately capture complex flows of scalar
transport phenomena with a combination of commodity hardware and 
powerful physics-constrained reconstructions. 
Our reconstruction method with its inflow solver and efficient tomography routine
are crucial for achieving this goal. The resulting algorithm allowed us
to assemble a first large-scale data set of realistic flows.
Our data set contains a unique combination of volumetric
data for turbulent flows in conjunction with visual data and processing
algorithms. In addition, our perceptual application demonstrated the usefulness
of the reconstructed data and led to first insights regarding smoke simulation
methods as well as the reconstruction itself. The studies show that the
advection-reflection solver as well as the wavelet turbulence model perform best
among the set of evaluated methods and our reconstructed flows contain
dynamics that are comparable to finely resolved simulations. 

The availability of a large, volumetric data set opens up a wide range of
possibilities. In particular, the availability of velocities in our
reconstructions means that the data can be flexibly used for re-simulations,
novel visualization, incorporation into VFX scenes, and metric
evaluations, e.g., in order to compare motions. Looking ahead, we believe that
there is a wide range of exciting future applications for our data.
Beyond avenues for benchmarking, accuracy measurements, and novel reconstruction
methods, we are looking forward to developments in the area of
machine learning that this data will enable \cite{sato2018example,xie2018tempogan,bkim2018deep}.

%%% Local Variables:
%%% TeX-master: "paper"
%%% End: